\documentclass[aps,prd,preprint,tightenlines,nofootinbib,showpacs,fixfloat]{revtex4}
\usepackage{epsfig}
\usepackage{graphicx}
\usepackage{dcolumn}
\usepackage{bm}
\usepackage{overpic}
\usepackage{subfigure}
\usepackage{float}

\usepackage{amsmath}
\usepackage{algpseudocode,algorithm,algorithmicx}  
\usepackage{amssymb}

\DeclareMathOperator*{\argmax}{arg\,max} 

\begin{document}

\title{Gaussian Process Accelerated Feldman-Cousins Approach for Physical Parameter Inference}
\author{Lingge Li, Nitish Nayak, Jianming Bian, Pierre Baldi}
\affiliation{%
 University of California, Irvine\\
}%
\date{\today}
             
\begin{abstract}
The unified approach of Feldman and Cousins allows for exact statistical inference of small signals that commonly arise in high energy physics. It has gained widespread use, for instance, in measurements of neutrino oscillation parameters in long-baseline experiments. However, the approach relies on the Neyman construction of the classical confidence interval and is computationally intensive as it is typically done in a grid-based fashion over the entire parameter space. In this letter, we propose an efficient algorithm for the Feldman-Cousins approach using Gaussian processes to construct confidence intervals iteratively. We show that in the neutrino oscillation context, one can obtain confidence intervals 5 times faster in one dimension and 10 times faster in two dimensions, while maintaining an accuracy above $99.5\%$. 
\end{abstract}

\maketitle

\section{Introduction}
Constructing classical confidence intervals for physical parameters with boundary conditions is challenging when dealing with small signals. The challenge is especially evident when studying neutrino oscillations because of the low event counts and multiple competing effects on the energy spectrum. 
The low event counts are primarily caused by the extremely low interaction cross-section of neutrinos, arising from the fact that they interact via the weak nuclear force. 
In order to extract meaningful statistical conclusions, one has to resort to means other than the asymptotic properties of Poisson data. The gold standard is the so-called unified approach outlined by Feldman and Cousins \cite{feldman1998unified}. It builds upon the Neyman construction of classical confidence intervals by specifying an ordering principle based on likelihood ratios and is known for providing correct coverage.
\par
The Feldman-Cousins approach is firmly grounded in statistical theory and widely used in neutrino experiments, for example Refs.~\cite{t2k}\cite{nova}\cite{minos}. However, it comes at a heavy computational cost, which in some cases such as Ref.~\cite{t2k} renders it infeasible for multi-dimensional confidence intervals. For the $1-\alpha$ confidence interval, the Feldman-Cousins approach includes all the values in the parameter space where the likelihood ratio test fails to reject at $\alpha$ level. However, it doesnʼt provide a prescription for how to sample that parameter space. Therefore, one is forced to sample it in its entirety in a grid-based fashion. Moreover, at each point one has to perform a large number of Monte Carlo simulations in order to calculate the $p$-value for the likelihood ratio test. 
\par
To accelerate the Feldman-Cousins approach, we propose approximating the function of $p$-values over the parameter space with Gaussian processes. Instead of performing a large number of Monte Carlo simulations, we start with just a small number of them at several parameter values to get noisy estimates of the $p$-values. We then train a Gaussian process model to interpolate over these estimates. Iteratively, we perform more Monte Carlo simulations to refine the Gaussian process approximation. We can control the $p$-value approximation error so that it does not change the likelihood ratio test decisions and the confidence interval. Meanwhile, the Monte Carlo simulations can be allocated intelligently in the parameter space to achieve substantial savings in computation.  
\par
The proposed algorithm is rooted in the framework of Bayesian optimization \cite{movckus1975bayesian}. It was originally designed to find the extremal points of an objective function that is unknown \textit{a priori}. In the Feldman-Cousins approach, the function of $p$-values over the parameter space is unknown. We adapt Bayesian optimization to locate a set of points in the parameter space that lie on the boundary of desired confidence intervals. By side-stepping points that are estimated to be either inside or outside the confidence interval with high probability, we can thus reduce the computational cost while producing the same result. We show that in the context of neutrino oscillation experiments, one can accelerate the construction of one-dimensional and two-dimensional confidence intervals by a factor of 5 and 10 respectively, without sacrificing the accuracy of the Feldman-Cousins approach. 


\section{Statistical Inference for Neutrino Oscillations}

\subsection{Neutrino Oscillations}
Neutrino oscillations demonstrate that neutrinos have mass and that the neutrino mass eigenstates are different from their flavor eigenstates. In the three flavor framework, the transformation of the mass eigenstates ($\nu_1$, $\nu_2$, $\nu_3$) into the flavor eigenstates ($\nu_e$, $\nu_\mu$, $\nu_\tau$) is described by the $3\times3$ unitary matrix $U_{PMNS}$ \cite{pontecorvo1957mesonium}, which is parameterized by three mixing angles $\theta_{12}$, $\theta_{23}$ and $\theta_{13}$, and a CP violation phase $\delta_{CP}$. The probability of oscillations between different neutrino flavor states of given energy $E_\nu$ over a propagation distance (baseline) $L$ depends on the $U_{PMNS}$ parameters and the difference of the squared masses of the eigenstates, $\Delta m^{2}_{32}$ and $\Delta m^{2}_{21}$.



The mixing angles $\theta_{12}$ and $\theta_{13}$ along with the squared-mass splitting $\Delta m^{2}_{12}$ have been measured to relatively high accuracy by several experiments, for example, Refs.~\cite{sno}\cite{sk}\cite{dayabay}. One can then infer the remaining parameters, $\theta_{23}$, $\delta_{CP}$, and $\Delta m^{2}_{32}$, by measuring the probabilities $P(\nu_{\mu} \rightarrow \nu_{\mu})$ and $P(\nu_{\mu} \rightarrow \nu_{e})$. Of particular interest are: (1) the sign of $\Delta m^{2}_{32}$, positive indicating a ``Normal Hierarchy" (NH) and negative indicating a ``Inverted Hierarchy" (IH) of neutrino mass states; (2) whether $\delta_{CP} \neq 0, \pi$, indicating Charge-Parity (CP) violation in the lepton sector; (3) whether the mixing angle is in fact maximal, i.e $\theta_{23} = 45^{\circ}$. The neutrino mass hierarchy has important implications for current and future neutrino experiments \cite{drexlin2013current} involved in measuring the absolute neutrino mass and investigating the possible Majorana nature of the neutrino. Leptonic CP-violation could be important to deduce the origin of the predominance of matter in the universe. 

\begin{figure}[h!]
  \centering
  \includegraphics[width=.45\linewidth]{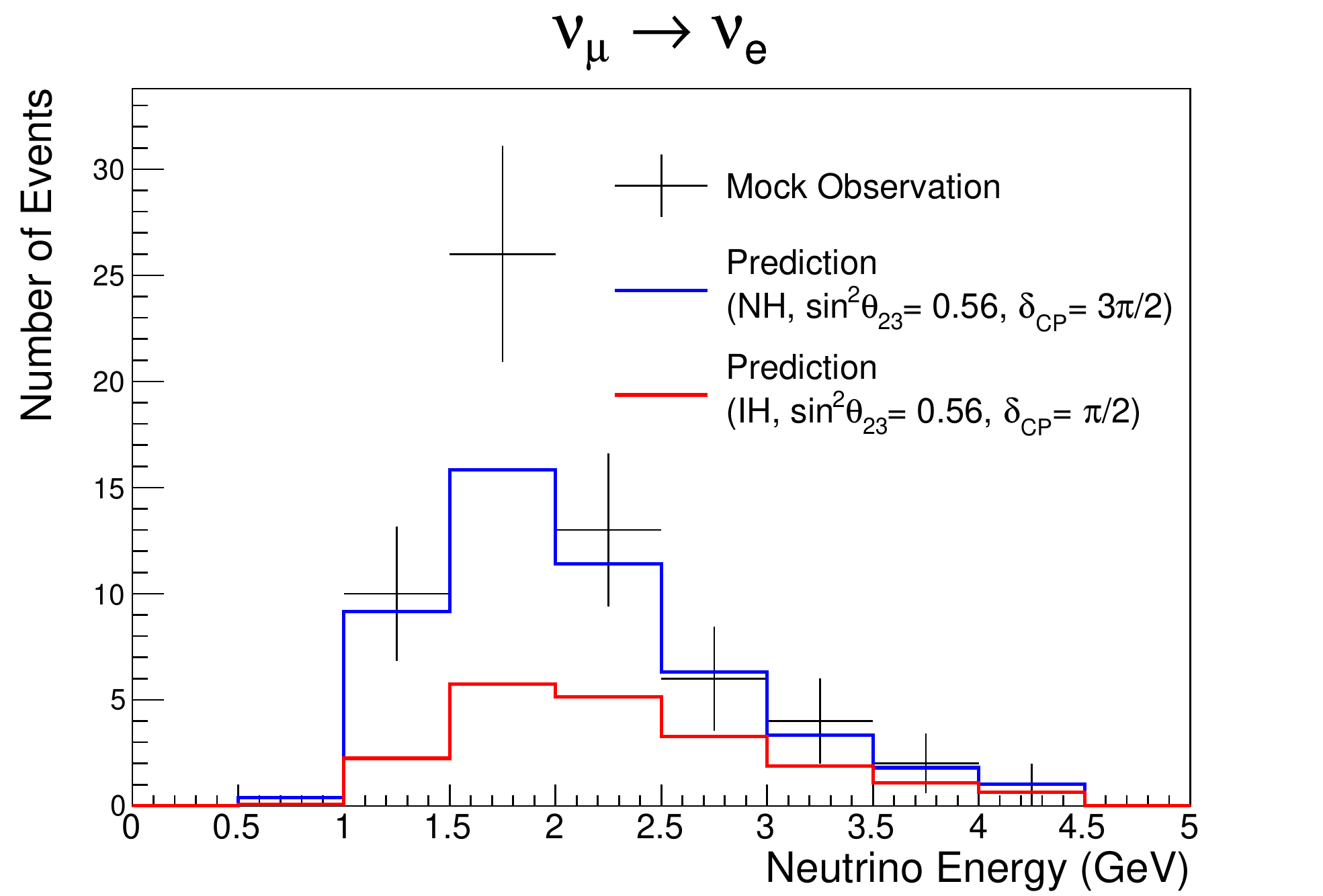}
  \includegraphics[width=.45\linewidth]{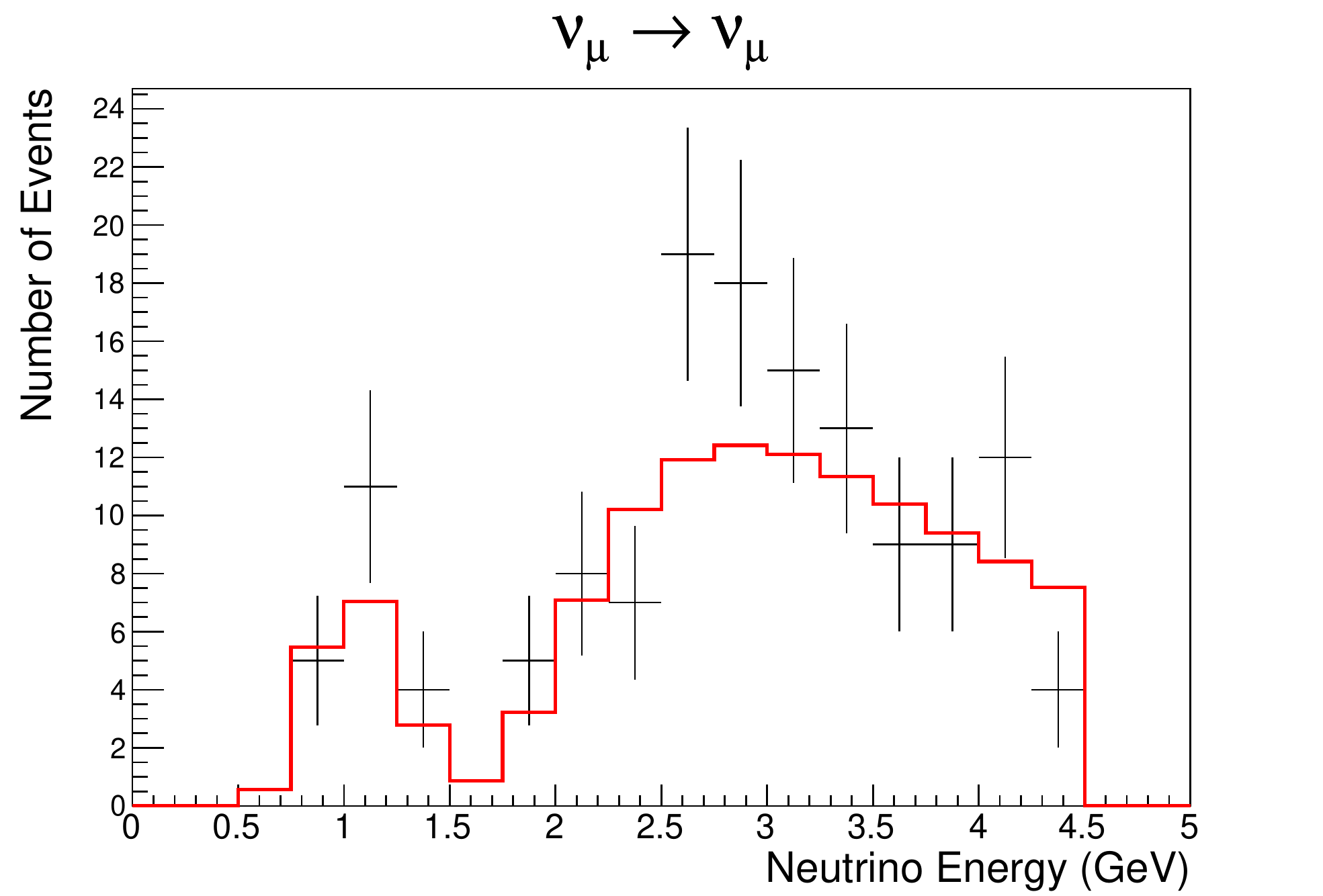}
\caption{An illustration of a toy neutrino oscillation experiment setup with the $\nu_{\mu} \rightarrow \nu_e$ channel on the left and the $\nu_{\mu} \rightarrow \nu_{\mu}$ on the right. Expectations for different oscillation parameters are compared to mock observations in order to find maximum likelihood estimates. The likelihood of observed data is maximized using the extended likelihood function. The fit is performed in both channels simultaneously.}
  \label{fig:spectrum}
\end{figure}

To infer neutrino oscillation parameters $\theta$, a typical long baseline neutrino oscillation experiment sends a beam of $\nu_\mu$ neutrinos into a detector and observes a handful of oscillated $\nu_e$ neutrinos along with $\nu_\mu$ neutrinos that survive over the baseline. As the oscillation probability is a function of neutrino energy, the observed neutrinos are binned by their energy. 
The neutrino oscillation parameters are inferred by comparing the observed neutrino energy spectra with the expected spectra for different oscillation parameters as shown in Fig.~\ref{fig:spectrum}.

\subsection{Feldman-Cousins Approach}
Denote the random variable for the neutrino count in the $i$-th energy bin by $X_i$. Further, assume that each $X_i$ follows an independent Poisson distribution with mean $\lambda_i$. For a given $\theta$, the expectations $\vec{\lambda}$ are also influenced by systematic uncertainties in the beam configuration and the interaction model among others, which we parameterize by $\delta$. For given oscillation and nuisance parameters $(\theta, \delta)$, the expectations $\vec{\lambda}$ given $(\theta, \delta)$ are obtained through simulations as they are analytically intractable. Denote the implicit mapping between $\vec{\lambda}$ and $(\theta,\delta)$ by $v$. The extended log-likelihood of $(\theta,\delta)$ is given by:
\begin{align*}
    \log L(\theta, \delta)&=\sum_{i\in I} \log Pois(x_i;v(\theta, \delta)_i)+\log Pois(\sum_{i\in I} x_i;\sum_{i\in I}v(\theta, \delta)_i)-\frac{1}{2}\delta^2 \label{likelihood}
\end{align*}
where $-\frac{1}{2}\delta^2$ is a penalty term for systematic error \cite{ext-ll}. 
\par
For a unified treatment of constructing classical confidence intervals for both null and non-null observations, an ordering principle based on likelihood ratios was introduced by Feldman and Cousins in 1997. The unified approach provides correct coverage even at parameter boundaries and has the highest statistical power as a result of the Neyman-Pearson lemma. In essence, a particular parameter value $\theta_0$ is included in the $1-\alpha$ confidence interval if the likelihood ratio test fails to reject the null hypothesis $\theta=\theta_0$ at the $\alpha$ level. The likelihood ratio test statistic is given by 
\begin{align*}
-2\log \frac{L(\theta_0)}{\argmax_\theta L(\theta)}
\end{align*}
 and has an asymptotic $\chi^2$ distribution by Wilks' theorem.
 
 \begin{figure}[h!]
  \centering
  \includegraphics[width=.45\linewidth]{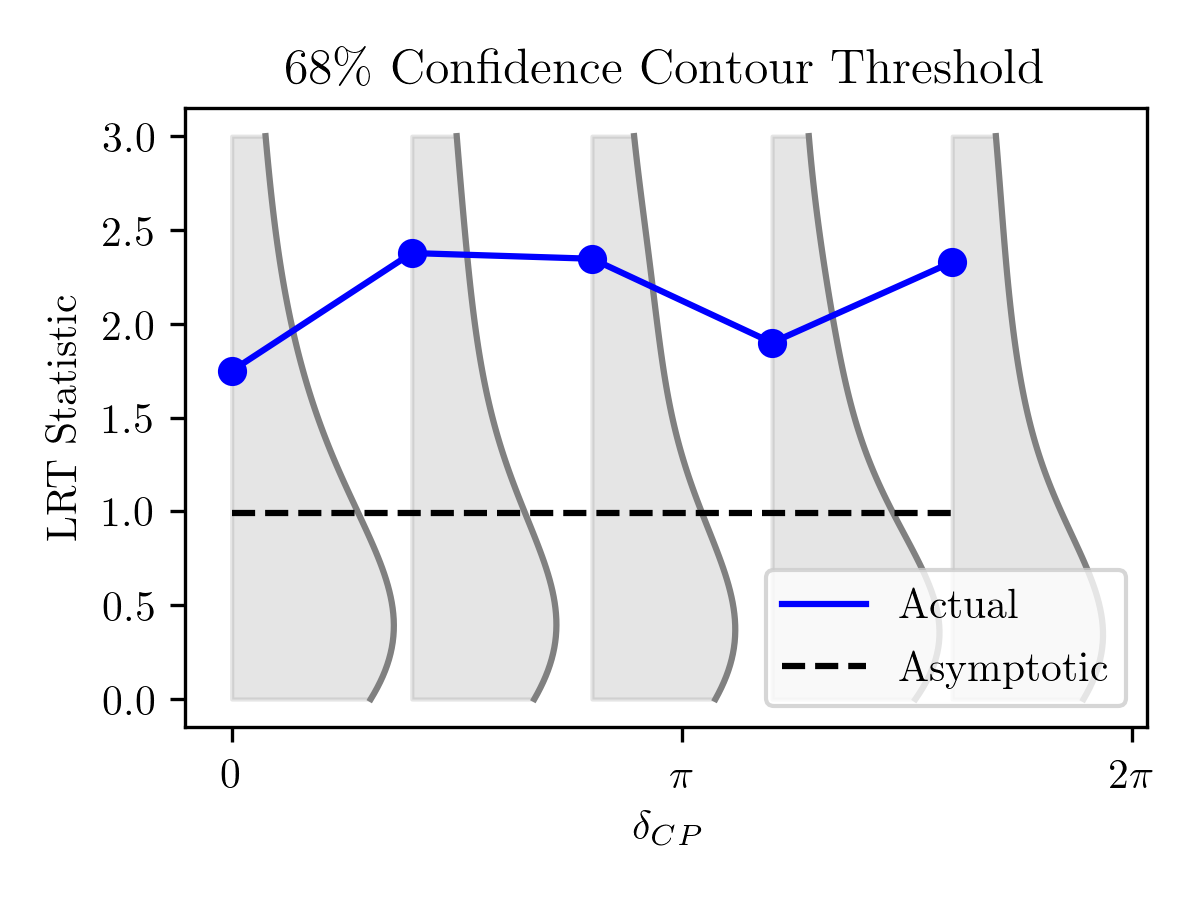}
  \caption{In the context of neutrino oscillations, the likelihood ratio test statistic distribution changes in the parameter space. Here the parameter is $\delta_{CP}$ and ranges from 0 to $2\pi$. The solid blue line indicates the $68$th-percentile of Monte Carlo simulated distributions while the dashed black line is the $68$th-percentile of the asymptotic $\chi^2_1$ distribution.}
  \label{fig:distribution}
\end{figure}

 In the context of neutrino oscillations, the asymptotic distribution is unreliable because of the small sample size in neutrino data and physical boundaries on the oscillation parameters. The reference distribution of the likelihood ratio test statistic can vary drastically as a function of $\theta$; Fig.~\ref{fig:distribution} shows several distributions at different $\theta$ values and comparisons of their critical values in particular. Therefore, for any given $\theta$, Monte Carlo experiments are used to simulate the reference distribution and calculate the $p$-value for the likelihood ratio test. Since the parameter space is bounded, the simulations are performed on a grid for a large number of $\theta$ values and the computational cost adds up quickly.

\section{Gaussian Process Algorithm}

\subsection{Gaussian Process Regression}
A Gaussian process ($\mathcal{GP}$) is a stochastic process where any finite collection of points are jointly Gaussian with mean $\mu$ and covariance $\Sigma$. An interpretation of the $\mathcal{GP}$ is an infinite extension of multivariate Gaussian; a $\mathcal{GP}$ can be thought of as a distribution in the function space where each draw from the distribution is a curve. Typically, the zero mean $\mathcal{GP}$ is used for modeling but it is still impossible to specify an infinite-dimensional covariance matrix explicitly. Instead, we can parametrize a zero mean $\mathcal{GP}$ with a kernel function $\kappa$ that defines the pairwise covariance.  Let $f\sim\mathcal{GP}(0, \kappa(\cdot,\cdot))$. Then for any pair $x$ and $x'$ we have
\begin{align*}
\left(\begin{array}{c} f(x) \\ f(x') \end{array}\right)\sim\mathcal{N}(0, \left[\begin{array}{c c} \kappa(x,x) & \kappa(x,x') \\ \kappa(x,x') & \kappa(x',x') \end{array}\right]).
\end{align*}

Given a finite set of observed data $\vec{x}_{obs}$, we can write down the multivariate Gaussian likelihood in this fashion and maximize it through kernel parameters $\omega$. Conveniently, at a new point $x^*$ we can obtain the closed form predictive distribution:
\begin{align*}
f(x^*)|f(\vec{x}_{obs})\sim N(\kappa(x^*, \vec{x}_{obs})(\kappa(\vec{x}_{obs},\vec{x}_{obs}))^{-1}f(\vec{x}_{obs}),\\ 
\kappa(x^*,x^*)-\kappa(x^*,\vec{x}_{obs})(\kappa(\vec{x}_{obs},\vec{x}_{obs}))^{-1}\kappa(\vec{x}_{obs},x^*)).
\end{align*}

\begin{figure}[h!]
  \centering
  \includegraphics[width=.4\linewidth]{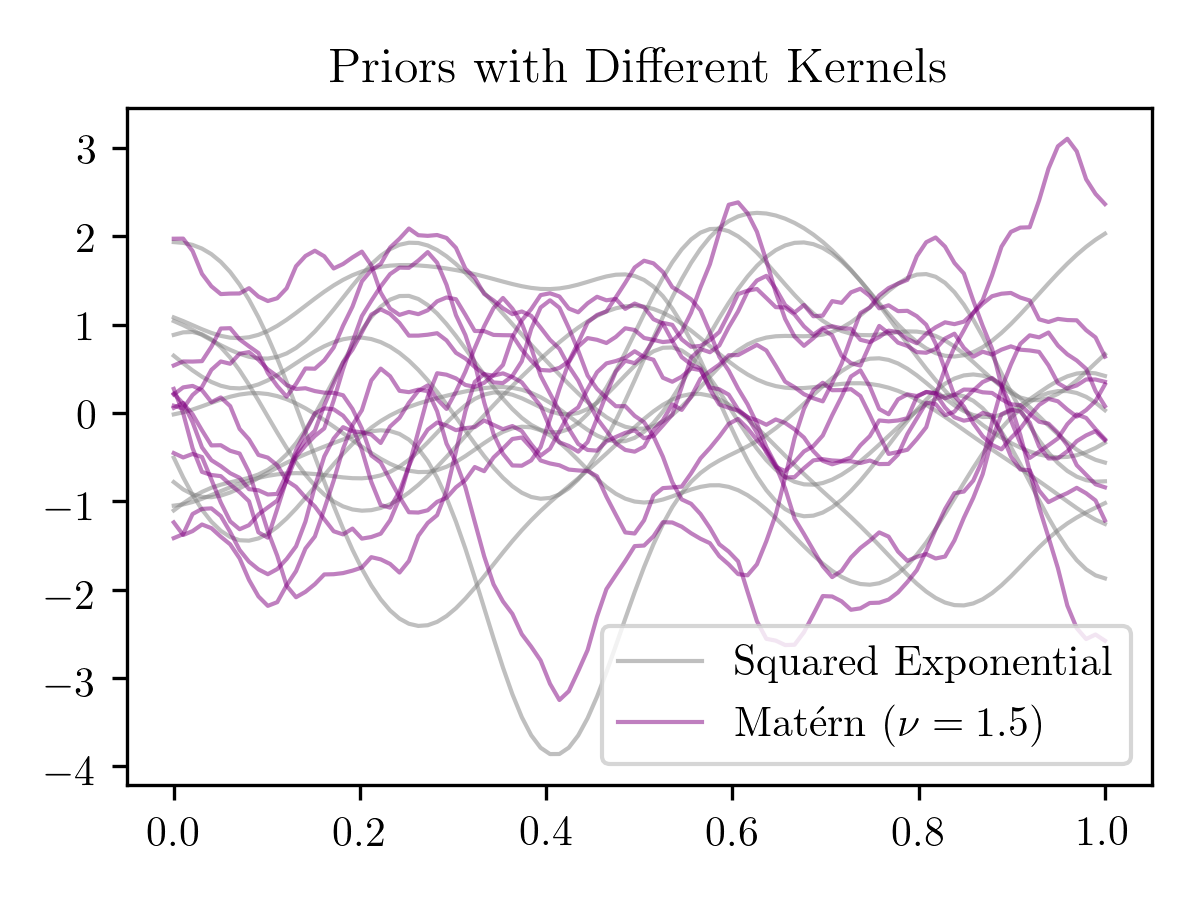}
  \includegraphics[width=.4\linewidth]{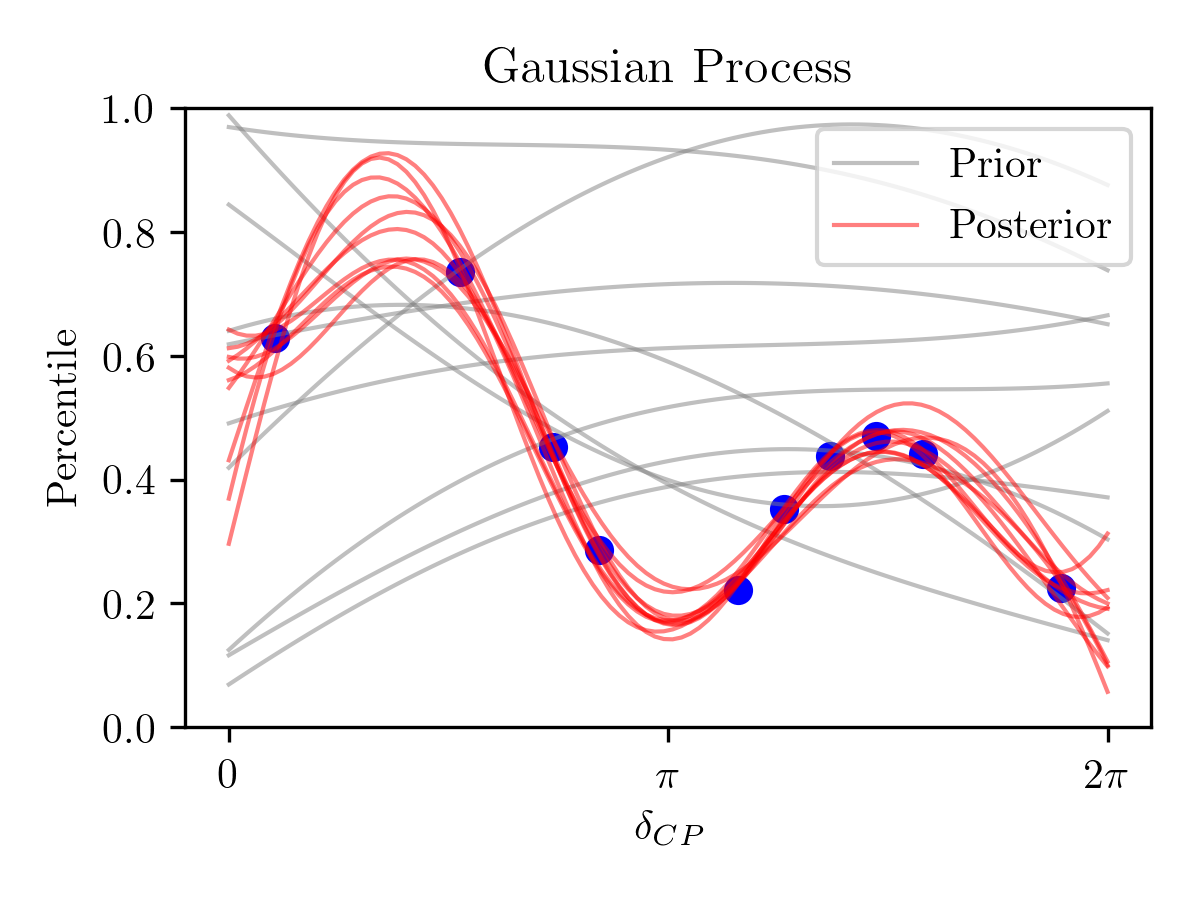}
  \caption{Sampled paths from Gaussian processes with different kernels (left). With $\nu=1.5$, the Mat\'{e}rn kernel produces functions that are only once differentiable. Sampled paths from Gaussian process prior and posterior with squared exponential kernel (right). The posterior paths, representing the curves drawn from the predictive distribution, are better aligned with the observed data points in solid blue.}
  \label{fig:gp}
\end{figure}

Since a $\mathcal{GP}$ is uniquely characterized by the kernel, different kernels produce distinct behaviors. A commonly used kernel is squared exponential $\kappa(x_1,x_2)=\exp(-(x_1-x_2)^2/l^2)$ where $l$ is called the length scale. Intuitively, the length scale determines the distance over which the $\mathcal{GP}$ interpolates between points. The squared exponential kernel is infinitely differentiable and functions drawn from such a $\mathcal{GP}$ would be smooth. However, this smoothness assumption might not be appropriate for some applications; a more general kernel is the Mat\'{e}rn kernel. The Mat\'{e}rn kernel has an additional parameter $\nu$ that controls the smoothness and the squared exponential kernel is a special case where $\nu\rightarrow\infty$. Fig.~\ref{fig:gp} shows some $\mathcal{GP}$ examples and please refer to Ref.~\cite{rasmussen2004gaussian} for more details on Gaussian processes.

Different kernels can be combined to compose a $\mathcal{GP}$ as long as the new kernel covariance matrix is still positive semi-definite. With the squared exponential kernel alone, the covariance implies that the observed data has no error. To account for error in the data, a diagonal matrix $\sigma^2 I$ is usually added to model constant variance across observations. In many situations such as ours, there exists heteroskedasticity, which means that different observations have different errors. When we iteratively perform Monte Carlo simulations to calculate $p$-values, the errors in the estimates also vary based on the number of simulations. We can actually model the $p$-value error as a diagonal matrix and add it to the $\mathcal{GP}$ covariance.

\subsection{Monte Carlo Error Estimation}

\begin{figure}[H]
  \centering
  \includegraphics[width=.4\linewidth]{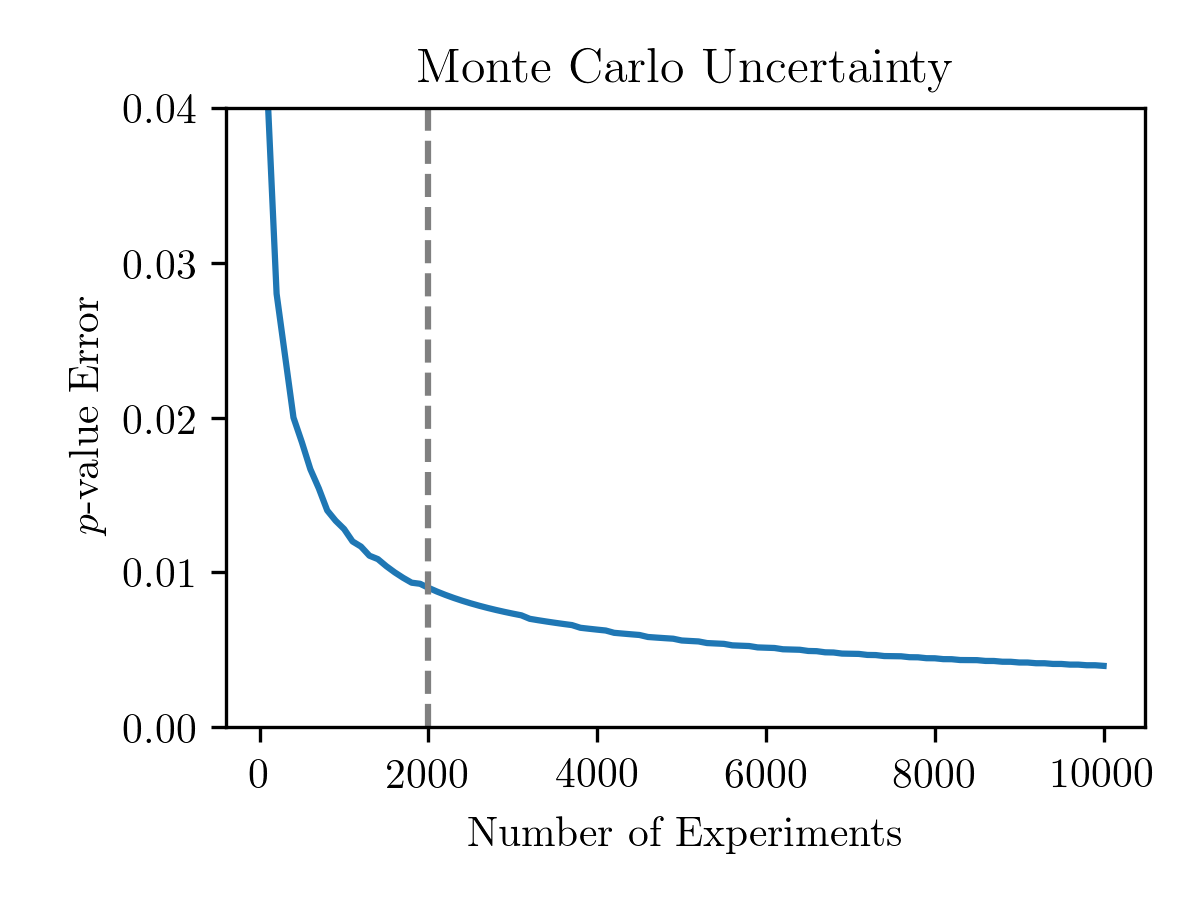}
  \includegraphics[width=.4\linewidth]{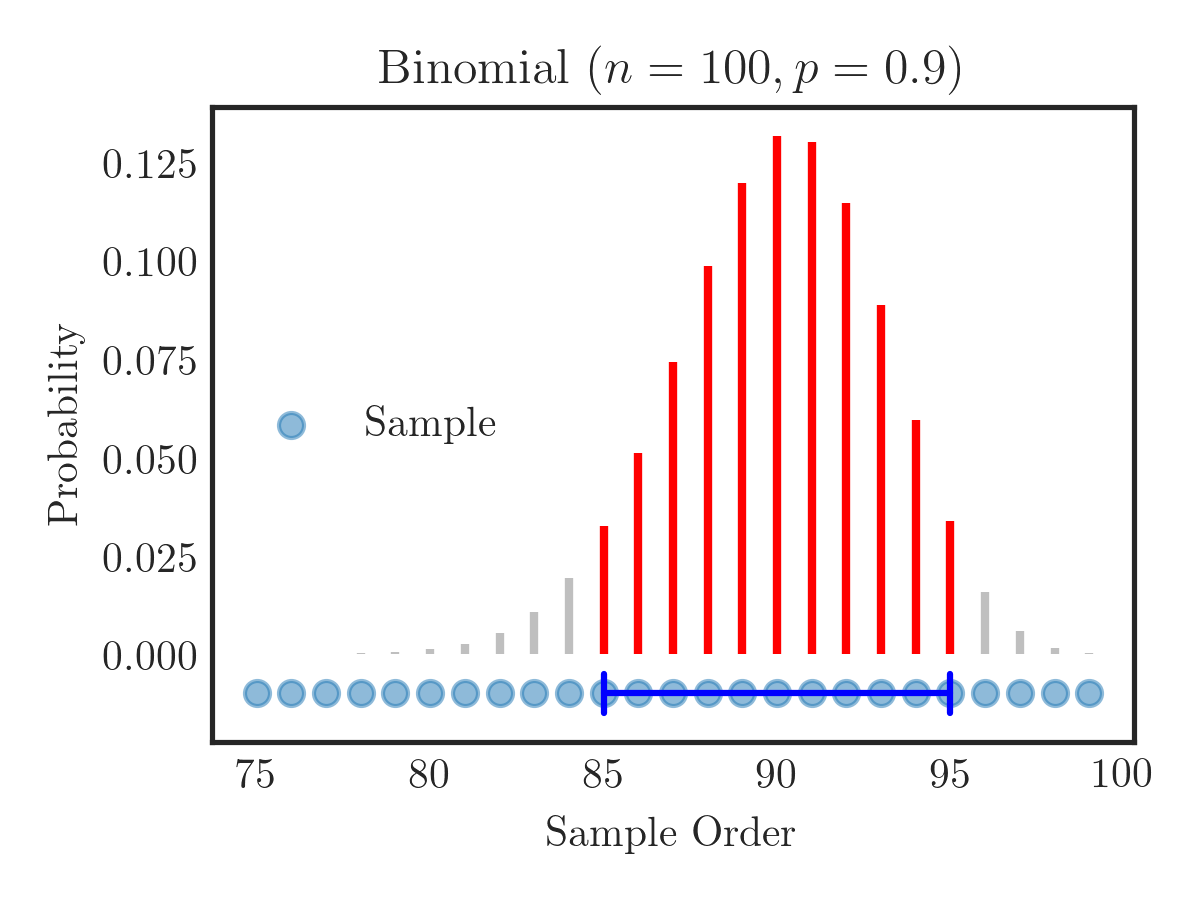}
  \caption{Monte Carlo error in terms of $p$-value as a function of the number of experiments (left). Example of non-parametric quantile interval construction using Binomial distribution (right). In a sample with 100 draws, the $85^{th}$ and $95^{th}$ order statistics form a 95\% confidence interval for the $90^{th}$ quantile of the unknown distribution.}
  \label{fig:variance}
\end{figure}

In the Feldman-Cousins approach, 
a large number of Monte Carlo simulations is required in order to make the error in $p$-value calculation negligible. When the Monte Carlo error in $p$-value calculation is not negligible, we should try to quantify it. Since the $p$-value is the quantile of the observed likelihood ratio statistic under the reference distribution, we can use a binomial proportion confidence interval as the $p$-value error estimate as outlined below \cite{hahn2011statistical}. As shown in Fig.~\ref{fig:variance}, the Monte Carlo error only slowly approaches zero when the number of simulations increases to 10,000. 

Suppose $X_1,...,X_n$ are independent draws from an unknown distribution $F$ whose $q^{\text{th}}$ quantile is denoted by $F^{-1}(q)$. Each draw $X_i$ is either below or above $F^{-1}(q)$ with probability $q$. Consequently, $M$, the number of $X_i$'s less than or equal to $F^{-1}(q)$, has a Binomial$(n,q)$ distribution. We can obtain a confidence interval for $F^{-1}(q)$ with sample statistics $X_{(l)},X_{(u)}$ (the $l^{\text{th}}$ and $u^{\text{th}}$ ordered draws) with $1\leq l\leq u\leq n$ such that
\begin{align*}
B(u-1;n,q)-B(l-1;n,q)\geq 1-\alpha.
\end{align*}
$B(u - 1;n,q) - B(l - 1;n,q)$ is the probability that $M$ is between $l$ and $u-1$. Thus, $(l,u)$ would form a confidence interval for $M$. Correspondingly, $(X_{(l)},X_{(u)})$ would form a confidence interval for $F^{-1}(q)$. 
Our goal, however, is to estimate $F(x^*)$ for an arbitrary $x^*$ given sample $X$, where, in our context, $x^*$ is the observed likelihood ratio test statistic and $F(x^*)$ is the $p$-value. This can be done by inverting the quantile confidence interval until the confidence intervals for $F^{-1}(q_l)$ and $F^{-1}(q_u)$ no longer contain $x^*$. Then $(q_l, q_u)$ would form a confidence interval for $F(x^*)$.

\subsection{Proposed Algorithm}
Bayesian optimization can be used to find the extremum of a black-box function $h$ when $h$ is expensive to evaluate so that a grid search is too computationally intensive. Bayesian optimization is an iterative procedure; in each iteration, $h$ is evaluated at a number of points to update an approximation of $h$. The approximation usually starts from a zero-mean Gaussian process prior $\mathcal{GP}(0,\kappa(\cdot,\cdot))$. After each iteration, the $\mathcal{GP}$ model yields a posterior distribution, hence Bayesian. Based on the approximation posterior, the points in the next iteration are proposed by an acquisition function $a$. The acquisition function $a$ aims to balance between ``exploration'', reducing approximation uncertainty, and ``exploitation'', reaching the extremum.

In our context, the expensive black-box function is the the function of $p$-values over the parameter space. 
Denote the grid points in the parameter space, where Monte Carlo simulations are performed, by $\vec{\theta}_{o}$, the simulated $p$-values at these points by $y(\vec{\theta}_{o})$, and the independent simulation errors by $\sigma(\vec{\theta}_o)$. The $\mathcal{GP}$ predictive posterior distribution of the unobserved $p$-values at $\vec{\theta}_u$ conditional on obtained $p$-values  $y(\vec{\theta}_{o})$ at points $\vec{\theta}_{o}$ is then given by 

\begin{align*}
f(\vec{\theta}_u)|y(\vec{\theta}_{o})\sim
\mathcal{N}(K_{uo}(K_{oo}+diag(\sigma^2(\vec{\theta}_o)))^{-1}y(\vec{\theta}_{o}), 
\\
K_{uu} - K_{uo}(K_{oo}+\sigma^2(\vec{\theta}_o)I)^{-1}K_{ou})
\end{align*}
where $K_{oo}, K_{ou}, K_{uo}, K_{uu}$ denote the covariance matrices between points $\vec{\theta}_o$ and $\vec{\theta}_u$.

\begin{figure}[h!]
  \centering
  \includegraphics[width=.45\linewidth]{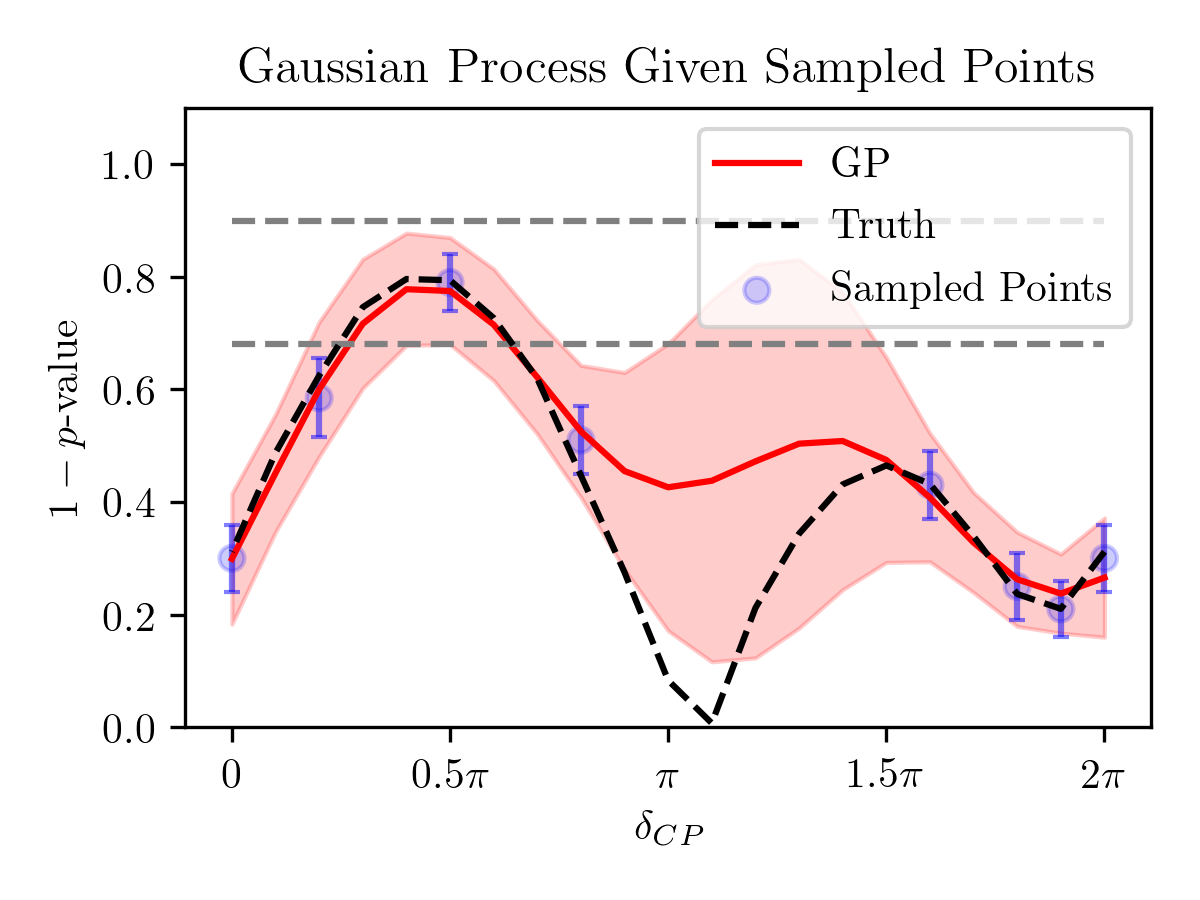}
  \includegraphics[width=.45\linewidth]{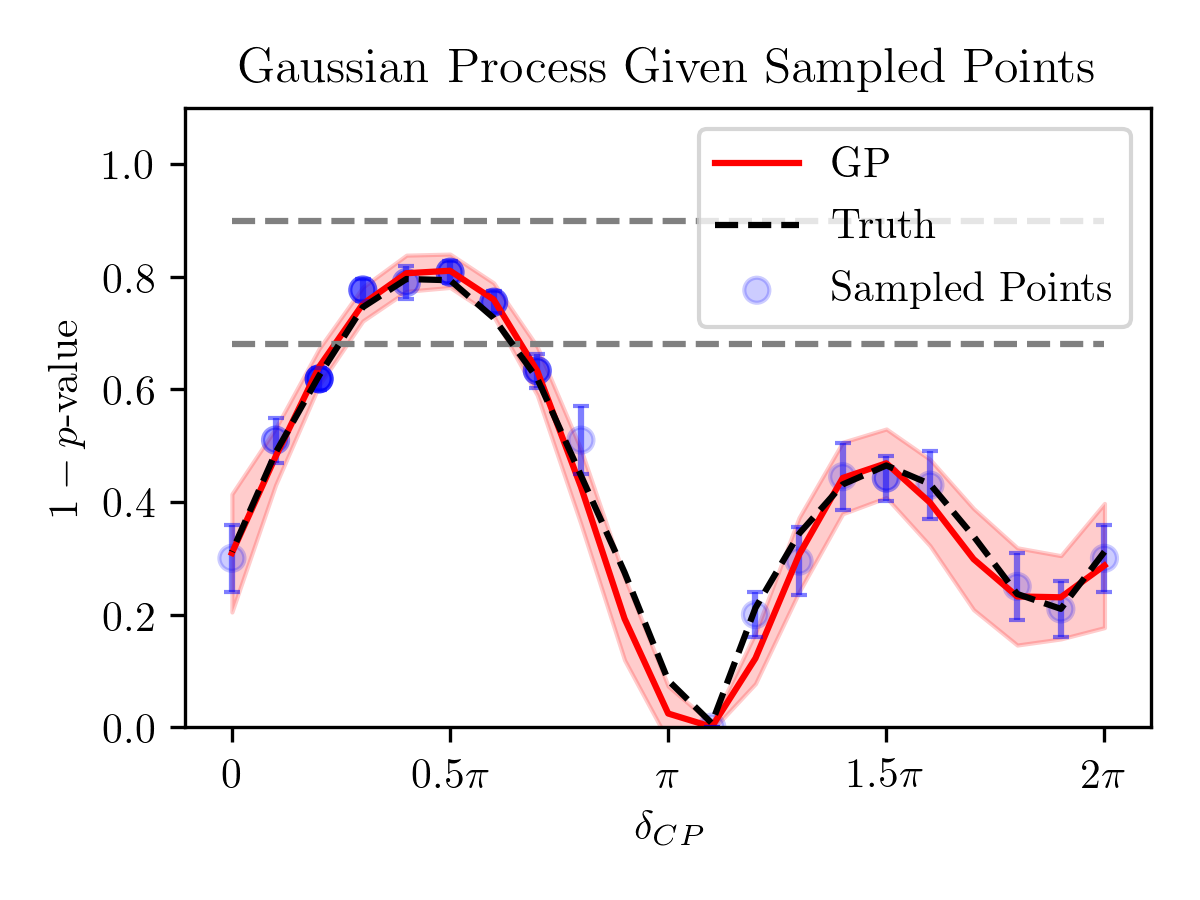}
  \caption{An illustration of our construction for the 68\% and 90\% confidence intervals for $\delta_{CP}$, which consist of points lying underneath the dashed horizontal lines. From a few initial points with high variance, the $\mathcal{GP}$ learns a rough approximation of the true curve (left). Based on the approximation, more points are proposed around the interval boundary, shown in dark blue, and the $\mathcal{GP}$ improves itself (right). The shade of blue represents the number of simulations used to calculate the $p$-value and the error bars are for the $p$-value.}
  \label{fig:example}
\end{figure}

Different from typical Bayesian optimization, we do not simply wish to find the minimum or maximum $p$-value. Instead, we want to find the points where the $p$-value is equal to $\alpha$ so that they enclose the confidence interval. Moreover, we want to be able to find multiple intervals at different confidence levels. Therefore, we choose our acquisition function to be
\begin{align*}
	a(\theta) = \sum_{\alpha_i}|\frac{f(\theta)-\alpha_i}{\sigma_{f(\theta)}}|^{-1}
\end{align*}
where $f(\theta)$ is the $\mathcal{GP}$ approximated $p$-value (posterior mean) at $\theta$ and $\sigma_{f(\theta)}$ is the $\mathcal{GP}$ posterior standard deviation at $\theta$. 
\begin{algorithm}[H]
\caption{$\mathcal{GP}$ iterative confidence interval construction}
\begin{algorithmic}
\For{each iteration $t = 1,2,...$}
\State Propose points in parameter space $\argmax_\theta a(\theta)$
\For{each point $\theta'$}
\State Simulate likelihood ratio statistic distribution
\For{$k = 1,2,...$}
\State Perform a pseudo experiment
\State Maximize the likelihood with respect to $(\theta, \delta)$
\State Maximize the likelihood with constraint $\theta=\theta'$
\State Calculate likelihood ratio statistic
\EndFor
\State Calculate $p$-value based on the simulated distribution
\EndFor
\State Train $\mathcal{GP}$ approximation $f(\theta)$ for the $p$-values
\State Update confidence intervals
\EndFor  
\end{algorithmic}
\end{algorithm}
Iteratively, the $\mathcal{GP}$ algorithm will seek points on the boundary of confidence intervals, for which it is unsure about. Points far from the boundary, which have $p$-values much greater or less than $\alpha_i$, are probabilistically ``ruled out.'' At these points, we will end up performing fewer Monte Carlo experiments or skipping them altogether. Every point on the grid would be either included or rejected with some uncertainty based on the $\mathcal{GP}$ posterior. With more iterations, the uncertainty will diminish so that the approximated confidence intervals converges to the ones produced by a full grid search. Fig.~\ref{fig:example} illustrates the proposed algorithm on an 1-dimensional example. 

Here we use the squared exponential kernel with the Monte Carlo errors added to the covariance diagonal and a small amount of white noise as often done in a regression setting \cite{rasmussen2004gaussian}. Point estimates of the $\mathcal{GP}$ kernel parameters by optimizing the log marginal likelihood
\begin{align*}
-\frac{1}{2}y(\vec{\theta}_{o})^T(K_{oo}+diag(\sigma^2(\vec{\theta}_o)))^{-1}y(\vec{\theta}_{o})-\frac{1}{2}\log |K_{oo}+diag(\sigma^2(\vec{\theta}_o))|-\frac{n}{2}\log 2\pi.
\end{align*}
There are constraints on the kernel parameters that should be incorporated. For instance, the length scale $l$ should be greater than the grid resolution and less than the grid range. 

\section{Numerical Studies}

\begin{figure}[H]
  \centering
  \includegraphics[width=.4\linewidth]{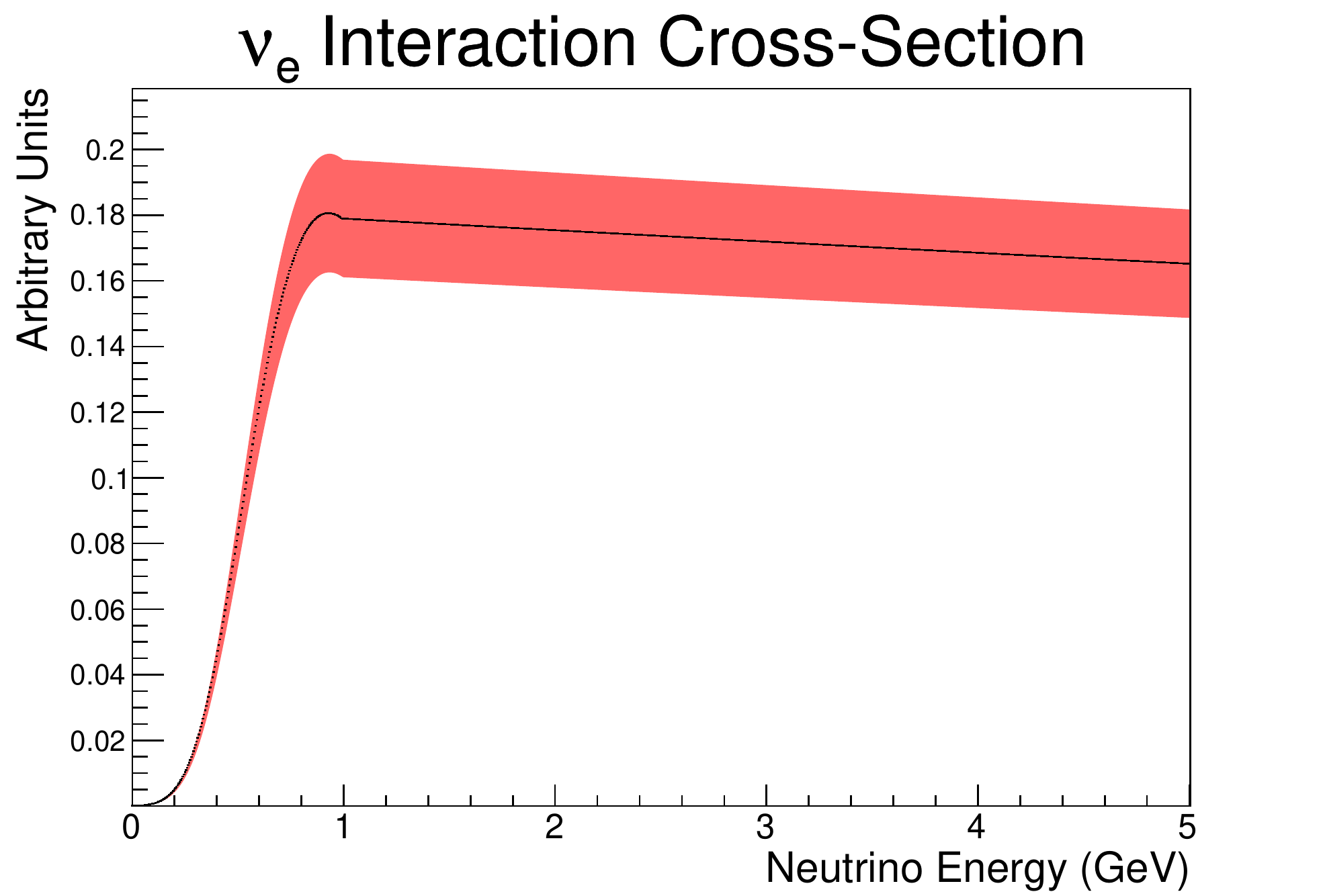}
  \includegraphics[width=.4\linewidth]{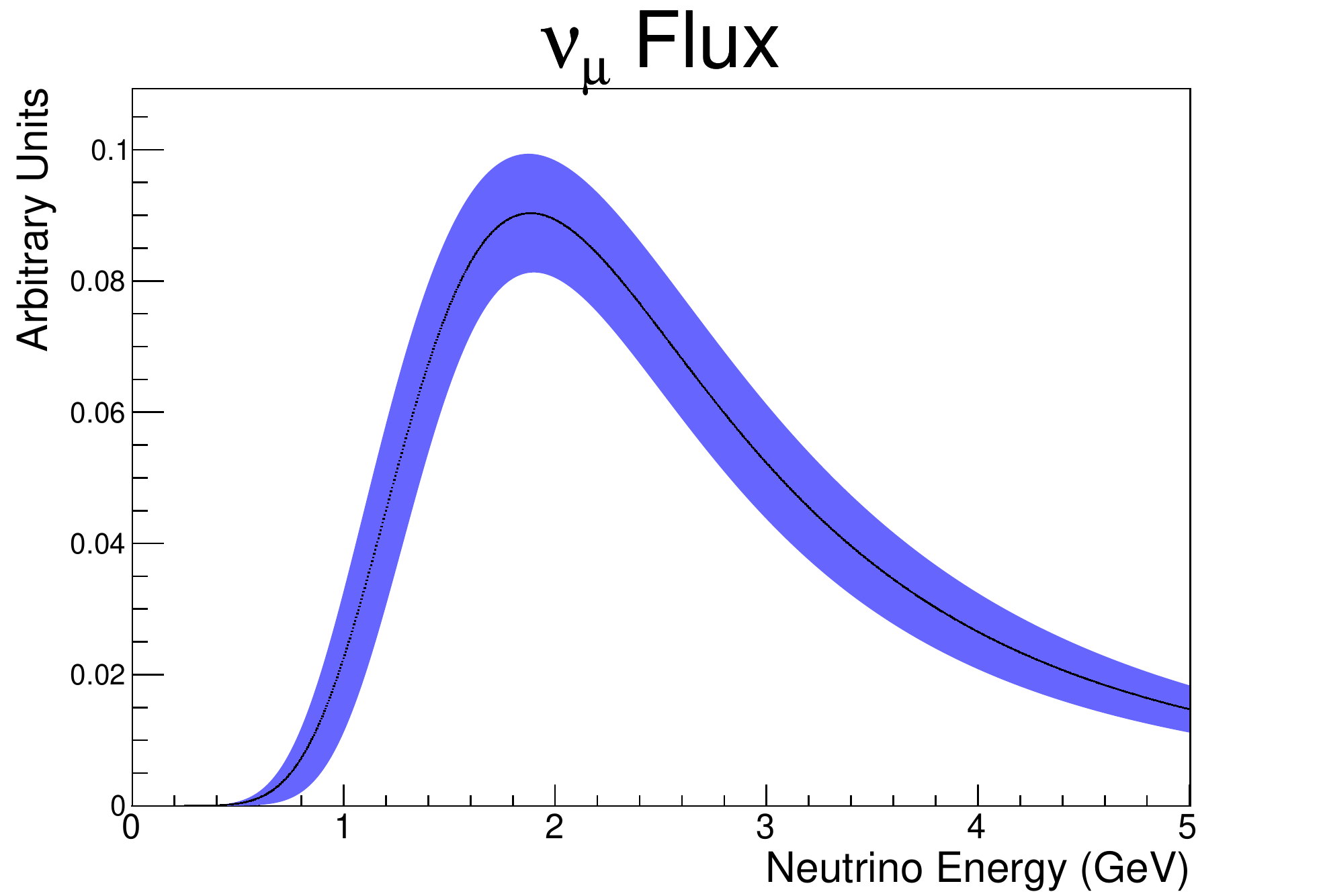}
  \caption{The distributions for $\nu_{e}$ interaction cross-section (left) and $\nu_{\mu}$ flux (right) are shown along with a normalization systematic error of 10\%.}
  \label{fig:toy}
\end{figure}

By way of illustration, we set up a toy long-baseline neutrino oscillation experiment in order to construct confidence intervals for the oscillation parameters. A flux distribution of $\nu_{\mu}$s is modeled as a Landau function over neutrino energies, $E_{\nu} \in (0.5, 4.5)$ GeV with the location parameter at $2$ GeV as shown in Fig~\ref{fig:toy}. The normalisation uncertainty is taken to be $10$\% and is applied as a nuisance parameter. The $\nu_{\mu}$ distribution is then oscillated into $\nu_{e}$s using the PMNS model for a toy baseline of $810$km through the Earth. Corrections from matter interactions \cite{MSW} are applied assuming a constant matter density of $2.84$ $g/cm^{3}$. The setup is similar to NOvA \cite{nova}, an accelerator-based long-baseline experiment at Fermilab. The oscillated $\nu_{e}$s are then ``observed" with a toy interaction cross-section distribution, similar in shape to Ref.~\cite{formaggio2012ev}; the cross-section increases as a function of neutrino energy from $0$ GeV up to $1$ GeV and decreases slowly until a maximum neutrino energy of $4.5$ GeV as shown in Fig~\ref{fig:toy}. A $10$\% normalisation uncertainty is applied on the cross-section as another nuisance parameter. Finally, we scale up the $\nu_e$ distribution to get an energy spectrum expectation, in energy bins of $0.5$ GeV between the flux range, similar to observations from NOvA \cite{nova}. The expected spectrum is computed from scratch for each set of oscillation and nuisance parameters in the toy experiment as shown in Fig.~\ref{fig:spectrum}. A similar setup is used for the $\nu_{\mu} \rightarrow \nu_{\mu}$ channel. However, in order to expedite the computation, the $2$-flavor oscillation probability approximation is used. The reactor mixing angle, $\theta_{13}$ and the solar parameters, $\theta_{12}$ and $\Delta m^{2}_{12}$ are fixed at the values given in Ref.~\cite{pdg}. A mock data set is obtained by applying Poisson variations on the expected spectrum at oscillation parameter values given by NOvA.

We then use this setup to construct 1-dimensional confidence intervals for $\delta_{CP}$ and 2-dimensional confidence intervals for $\sin^2\theta_{23}$ vs $\delta_{CP}$ by the two algorithms, a standard grid-search implementation of Feldman-Cousins and the $\mathcal{GP}$-based algorithm.  $|\Delta m^2_{32}|$ is treated as a nuisance parameter while $\sin^2 \theta_{23}$ is treated as another in the case of the 1-dimensional interval for $\delta_{CP}$. The likelihood function is integrated over the nuisance parameters assuming a flat prior in the range $(2, 3)$ $\times10^{-3}$ eV$^{2}$ for $|\Delta m^2_{32}|$ and $(0.3, 0.7)$ for $\sin^2 \theta_{23}$, similar to Ref.~\cite{t2k}. The prior on the nuisance parameters for the systematic uncertainties in the toy model is assumed to be a standard normal distribution. The toy model and parameter fitting routine are implemented in ROOT \cite{brun1997root} while the Gaussian process algorithm is implemented with scikit-learn \cite{pedregosa2011scikit}.

\subsection{1-dimensional Confidence Intervals}

\begin{figure}[H]
  \centering
  \includegraphics[width=.45\linewidth]{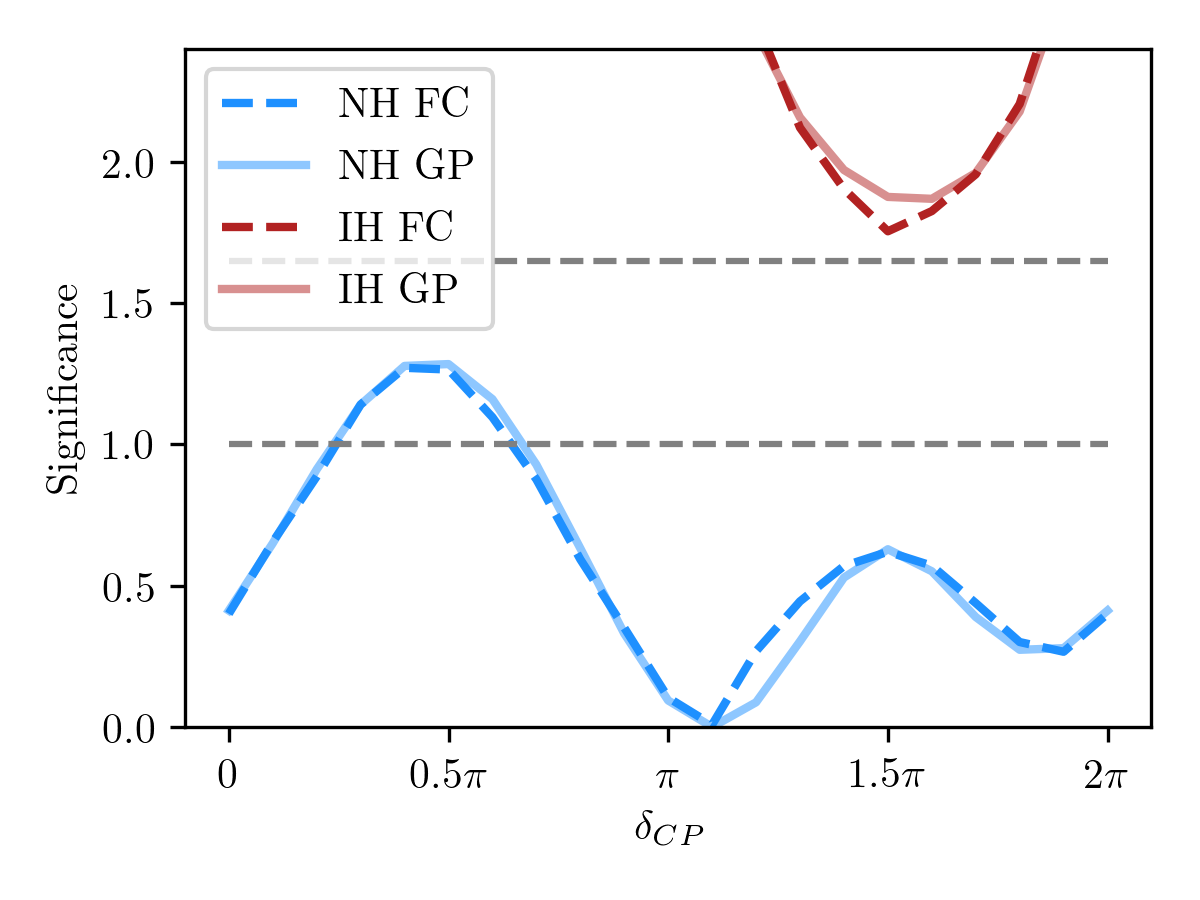}
  \caption{Example significance curves obtained with the standard Feldman-Cousins and Gaussian process algorithms mostly overlap, especially when the significance is close to $1\sigma$ and $1.6\sigma$ as desired. In this case, the inverted hierarchy (IH) is rejected at $1.6\sigma$ level and the normal hierarchy (NH) has the same $1\sigma$ confidence interval.}
  \label{fig:1d_curves}
\end{figure}

To make inference on $\delta_{CP}$, a significance curve is usually drawn under different assumptions of mass hierarchy as shown in Fig.~\ref{fig:1d_curves}. The portion of the significance curve below a certain value gives us the confidence interval at that level. We can observe that the NH curves by both the standard FC and $\mathcal{GP}$ algorithms have the same intersections with $1\sigma$ horizontal line, which implies that the $1\sigma$ confidence intervals are the same. 
Though there are slight discrepancies, the shape of the $\mathcal{GP}$ significance curve is mostly correct.

\begin{figure}[H]
  \centering
  \includegraphics[width=.45\linewidth]{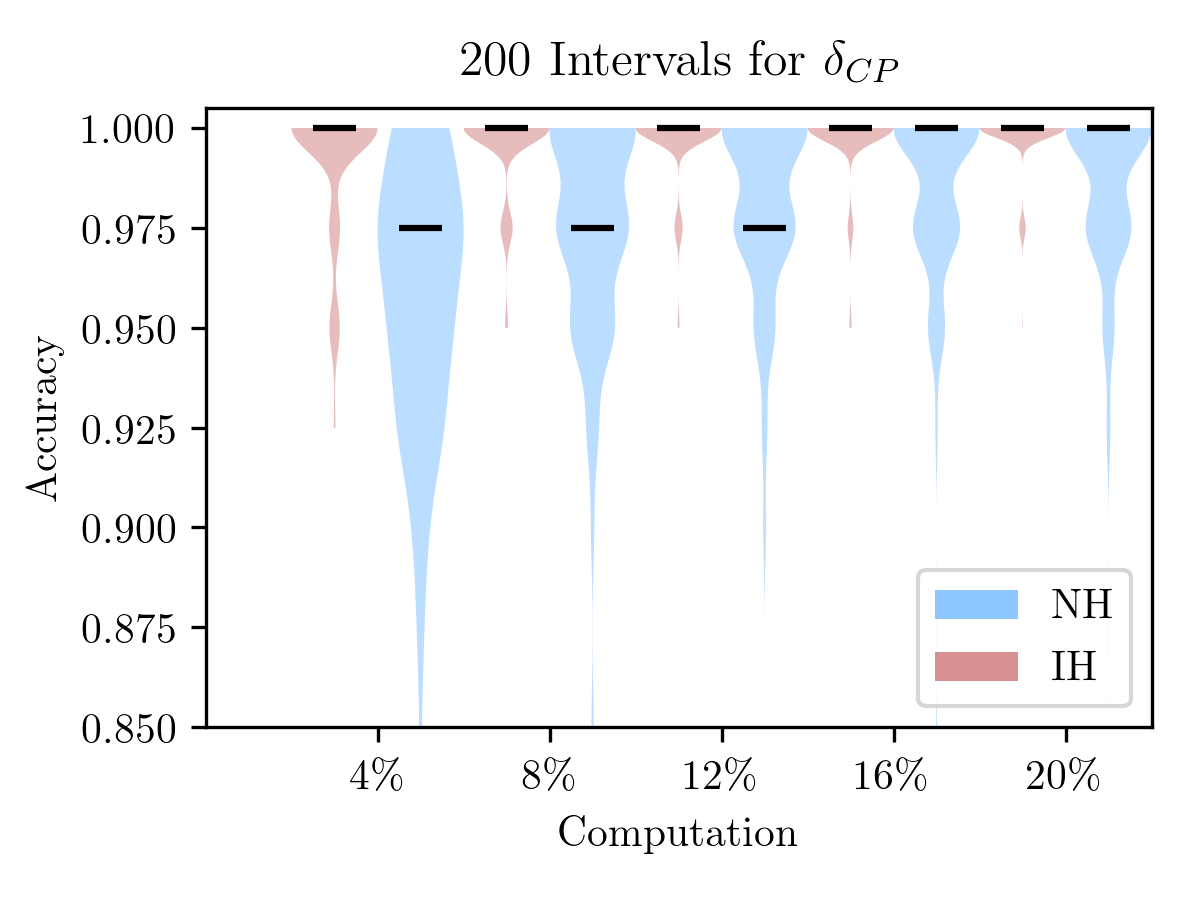}
  \includegraphics[width=.45\linewidth]{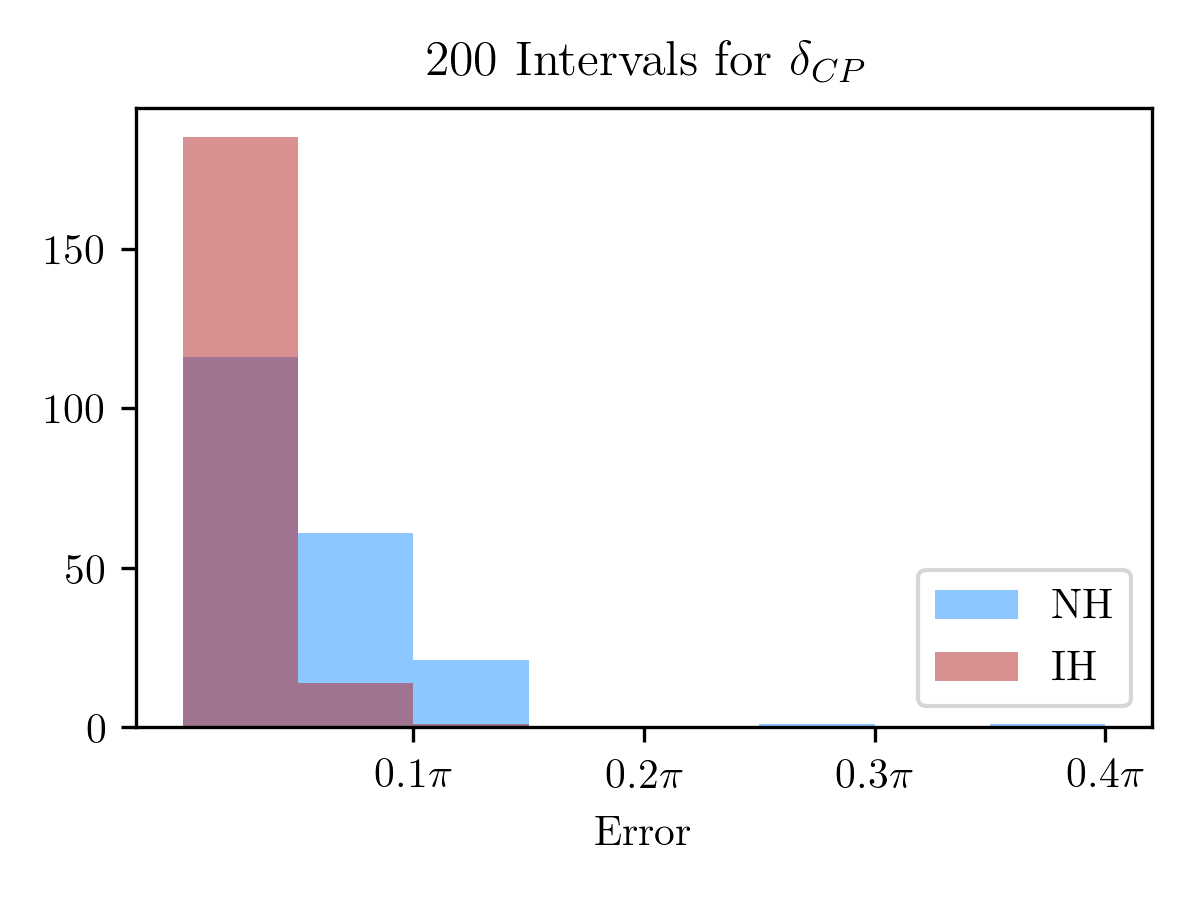}
  \caption{Relative accuracy of the confidence intervals in terms of correctly included grid points as a function of computation (left) and the distribution of absolute errors for both normal and inverted hierarchies (right).}
  \label{fig:1d_results}
\end{figure}

To evaluate the performance of the $\mathcal{GP}$ algorithm, we perform the same inference procedure on 200 different data sets to find the 68\% and 90\% confidence intervals. First, with standard FC results as ground truth, we consider the accuracy of the $\mathcal{GP}$ algorithm for classifying whether or not each grid point is included in the confidence intervals. As the $\mathcal{GP}$ algorithm is iterative, we can calculate the accuracy at the end of each iteration with fixed computation. When the computation reaches 20\% of that is required by standard FC, we stop the algorithm and calculate the absolute error as the difference in confidence interval endpoints. Fig~\ref{fig:1d_results} shows that the median accuracy reaches 1 with less than 20\% of computation and the error is no more than $0.1\pi$ for most data sets. As $\delta_{CP}$ ranges from $0$ to $2\pi$ and there are only 20 grid points, an error of $0.1\pi$ is just one grid point. With a finer grid, we expect the performance of the $\mathcal{GP}$ algorithm to improve.    

\subsection{2-dimensional Confidence Contours}

\begin{figure}[h!]
  \centering
  \includegraphics[width=.45\linewidth]{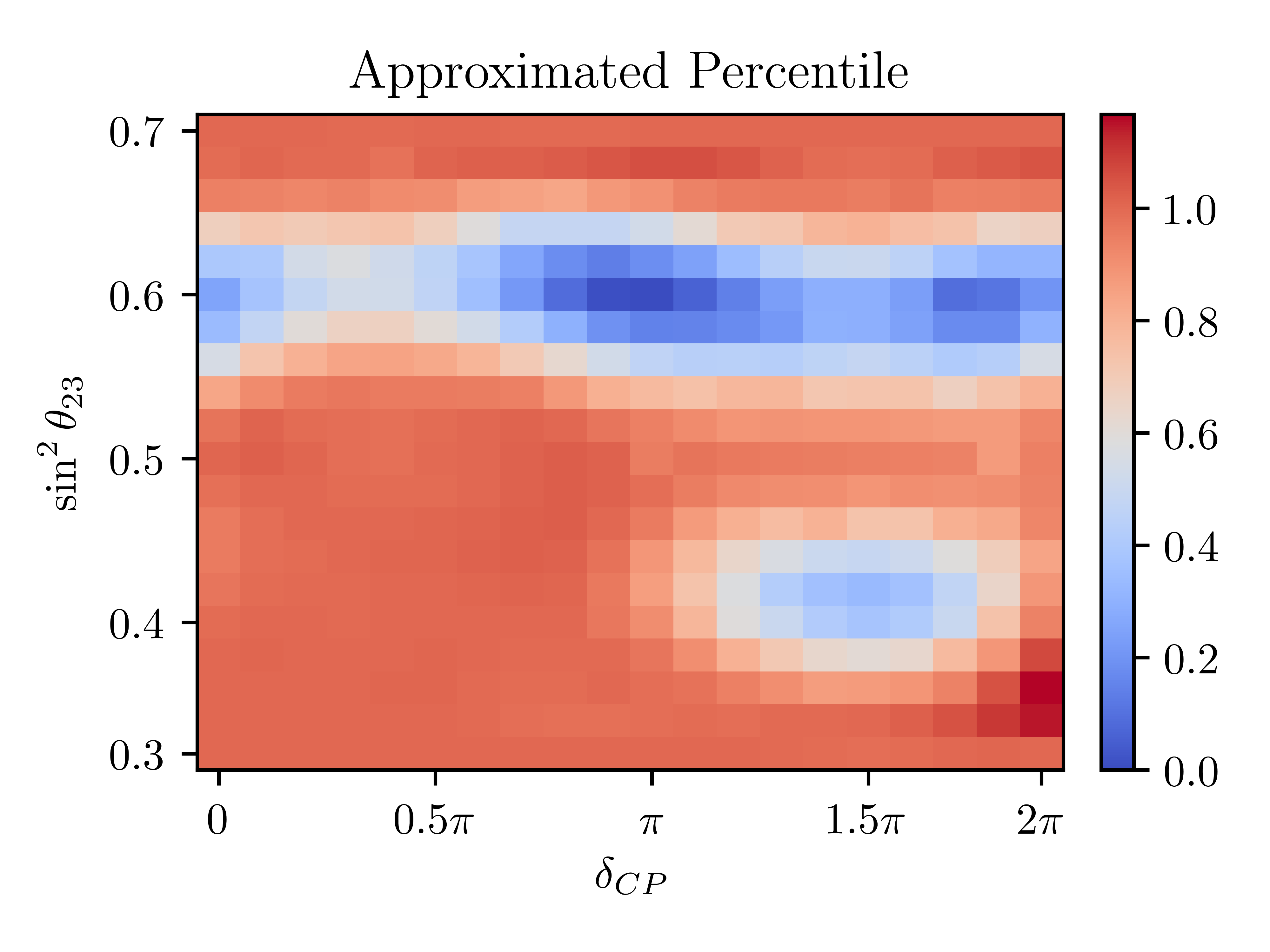}
  \includegraphics[width=.45\linewidth]{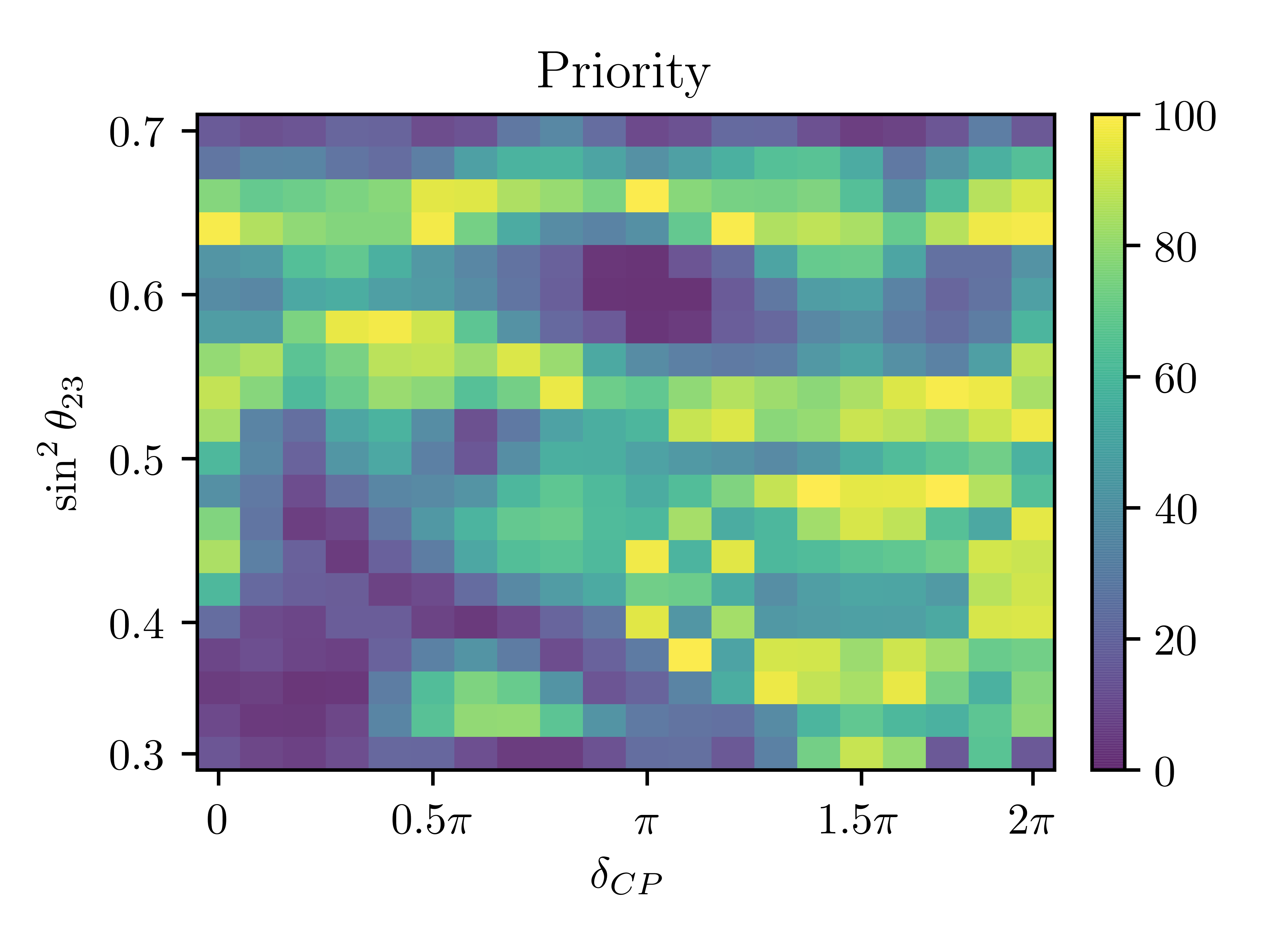}
  \caption{$\mathcal{GP}$ approximated percentile ($1-p$-value) on the $20\times20$ grid for $\sin^2 \theta_{23}$ vs $\delta_{CP}$ (left) and the priority to sample points from the grid (right). Notice that the points near $68\%$ and $90\%$ have the highest priority.}
  \label{fig:2d_iteration}
\end{figure}

\begin{figure}[h!]
  \centering
  \includegraphics[width=.45\linewidth]{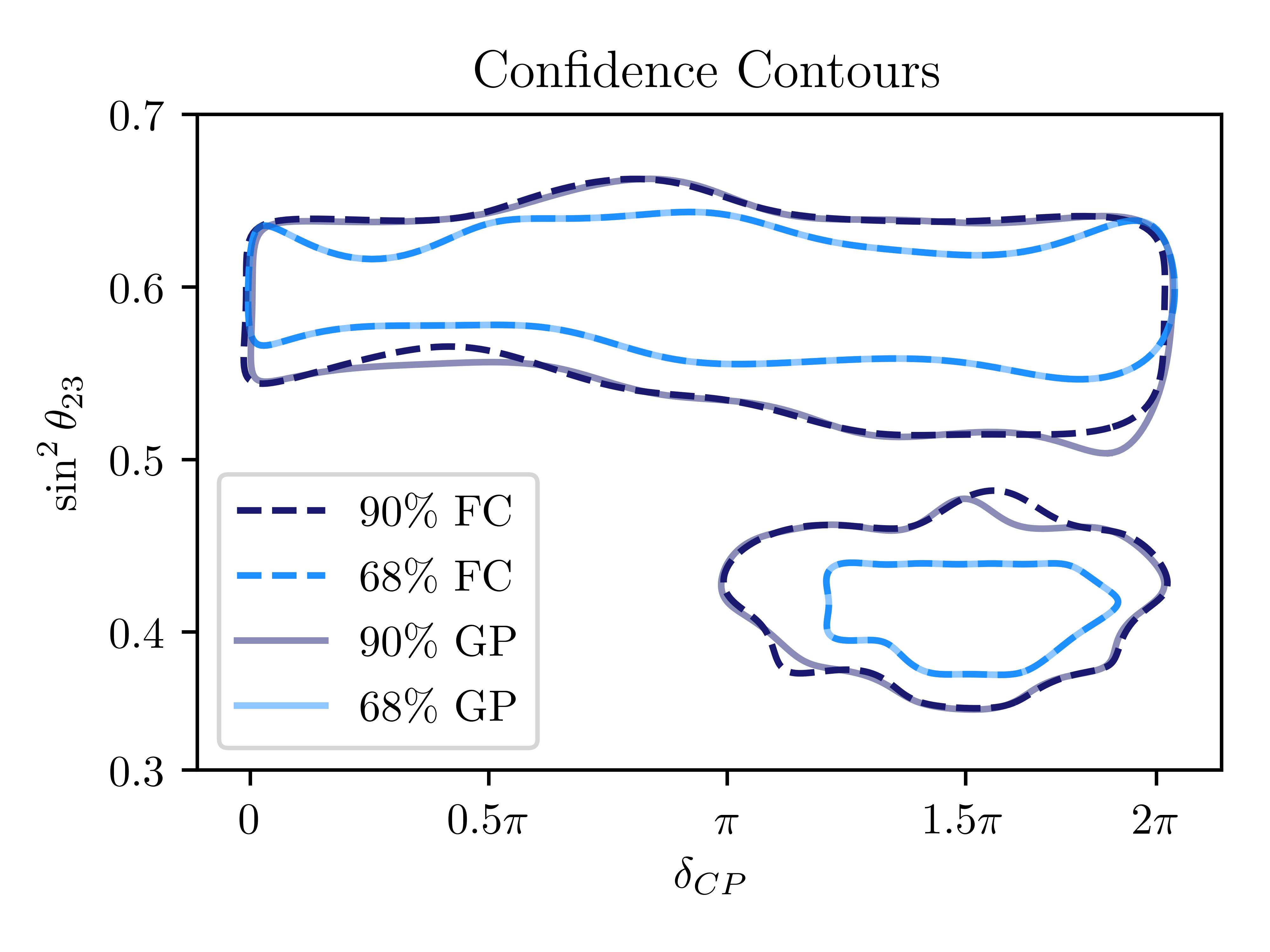}
  \includegraphics[width=.45\linewidth]{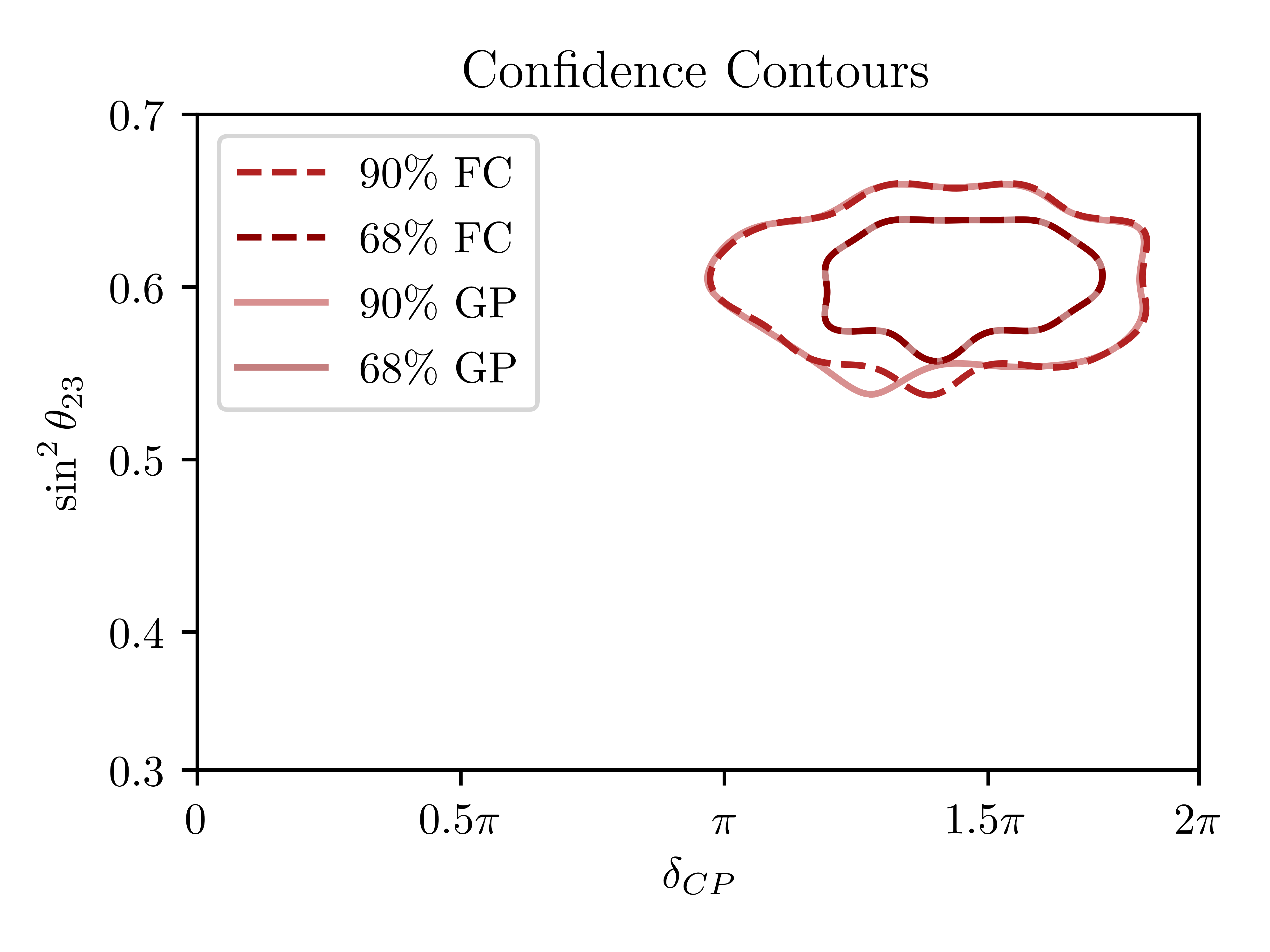}
  \caption{Confidence contours for the same data constrained to normal (left) and inverted hierarchies (right). The true (dashed) and approximated (transparent) contours are almost indistinguishable.}
  \label{fig:2d_contours}
\end{figure}

To find the 2-dimensional confidence contours under hierarchy constraints, the $\mathcal{GP}$ algorithm approximates the $p$-value surface on the parameter grid as shown in Fig.~\ref{fig:2d_iteration} and specifically prioritizes points on the contour boundaries. Grid points below a certain value are included in the confidence contour at that level. To make the final smooth contours in Fig.~\ref{fig:2d_contours}, we use Fourier smoothing to draw the closest elliptical curves. We can observe that the FC and $\mathcal{GP}$ contours overlap in the same areas. In fact, the area difference between the contours is on the same order of magnitude with Fourier smoothing. 

\begin{figure}[h!]
  \centering
  \includegraphics[width=.45\linewidth]{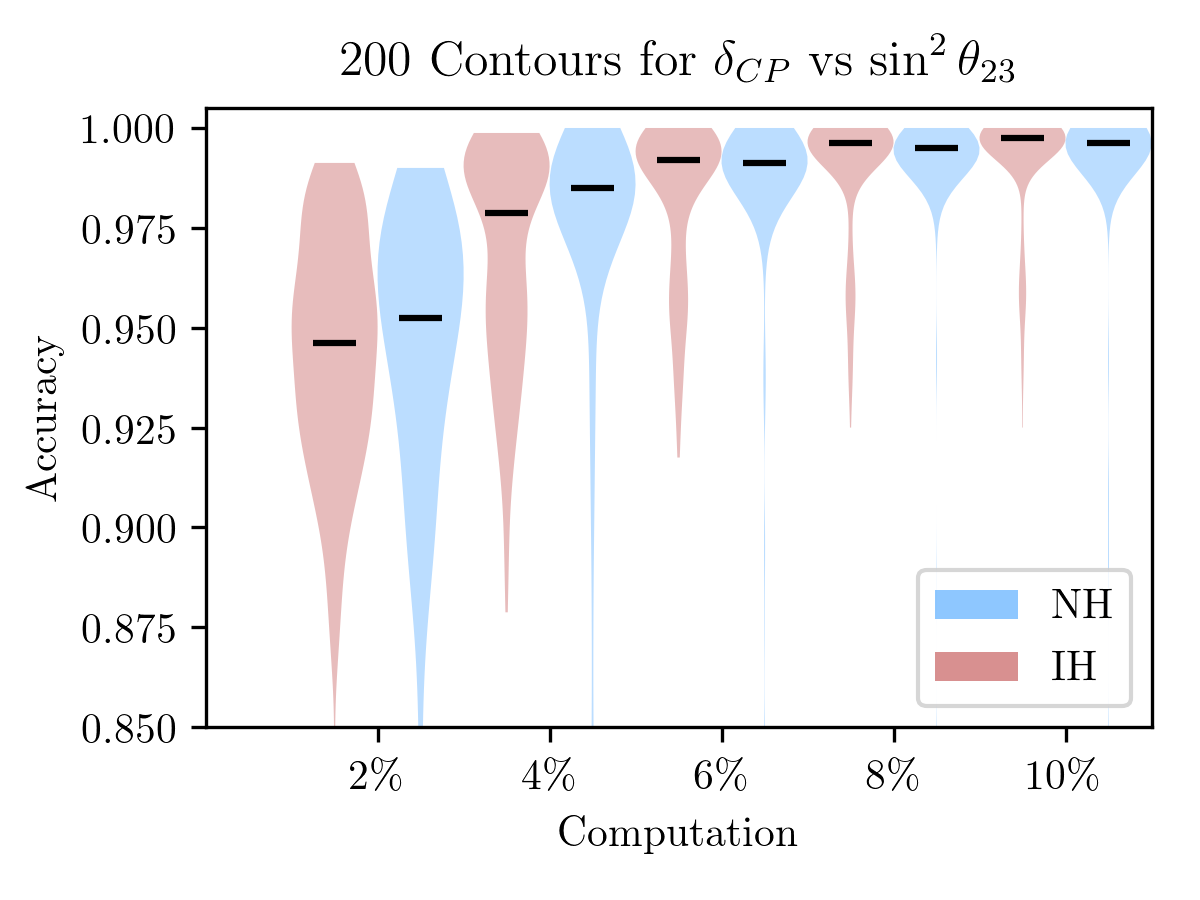}
  \includegraphics[width=.45\linewidth]{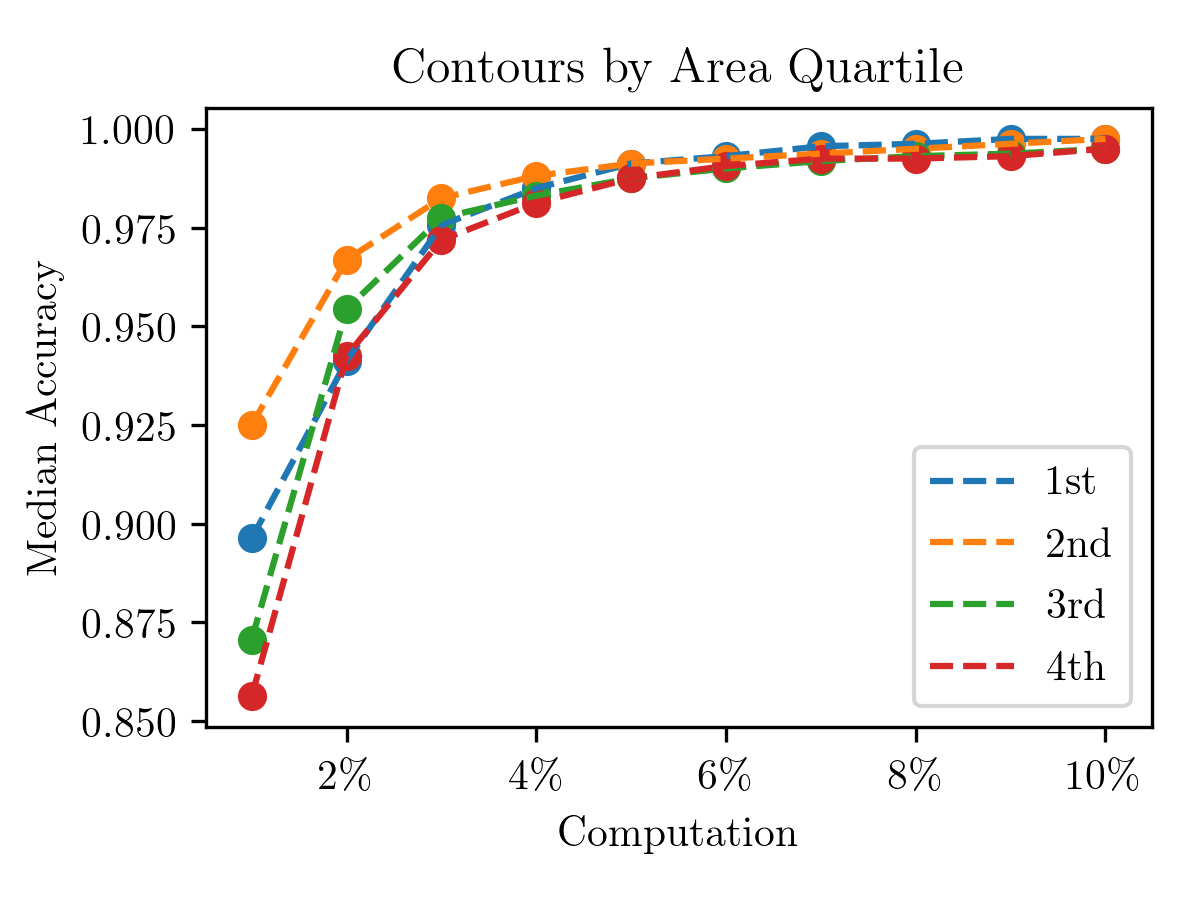}
  \caption{Relative accuracy of the confidence contours as a function of computation (left) and median accuracy stratified by area as a function of computation (right).}
  \label{fig:2d_results}
\end{figure}

Similarly, we use both algorithms on 200 different data sets to find the 68\% and 90\% confidence contours and calculate the grid point classification accuracy after each iteration up to 10\% of the standard FC computation. A concern is that contours with larger area could require more computation to achieve the same accuracy as there are more points along the boundary. We address this concern by stratifying contours by area quartile and plotting median accuracy as a function of computation. Fig.~\ref{fig:2d_results} shows that the median accuracy reaches 1 with less than 10\% of computation and contour area does not have an effect. The reason is that while larger contours have more points on the boundary, smaller contours are more difficult to locate precisely. Overall, it takes roughly the same computation to probe the $p$-value surface accurately so the $\mathcal{GP}$ algorithm should have similar performance regardless of the contour size. 

\begin{figure}[h!]
  \centering
  \includegraphics[width=.4\linewidth]{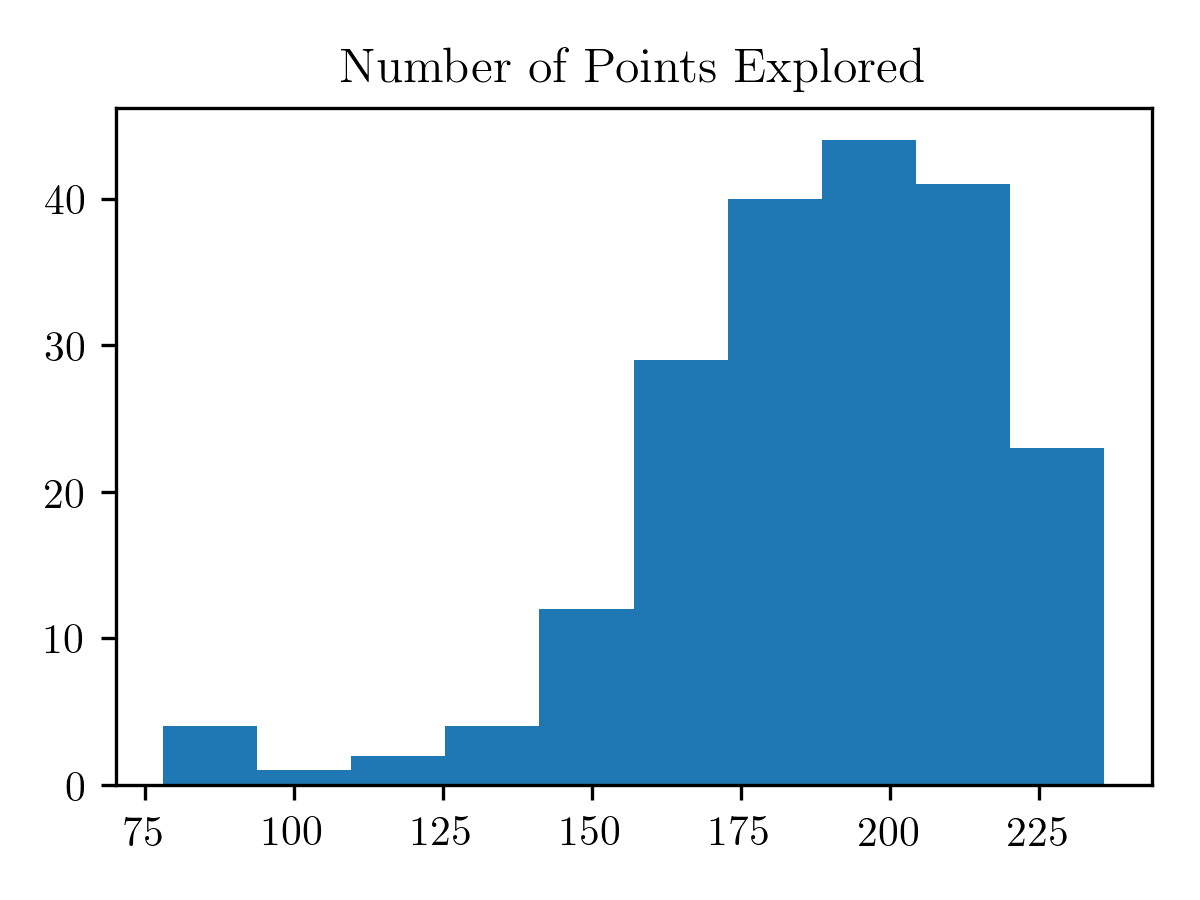}
  \includegraphics[width=.4\linewidth]{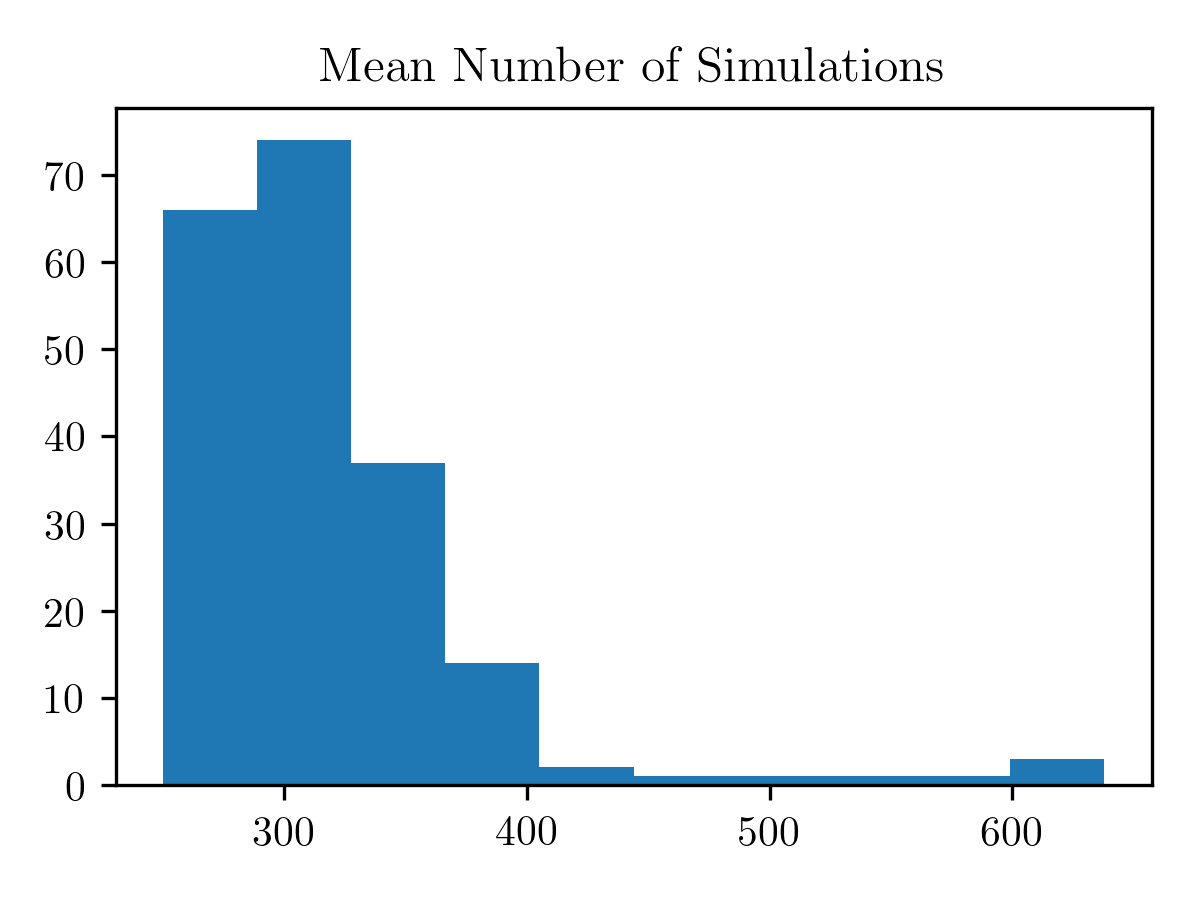}
  \caption{Distribution of the number of points explored on the grid (left) and distribution of the average number of Monte Carlo experiments simulated at a point (right).}
  \label{fig:2d_savings}
\end{figure}

Lastly, we are interested in where the computational savings come from. We keep track of the number of grid points explored by the $\mathcal{GP}$ algorithm and the number of simulations at each point for the 200 data sets. Fig.~\ref{fig:2d_savings} shows that the algorithm explores about half of the total grid points and on average only about 300 Monte Carlo simulations are done instead of 2000 in standard FC. We conclude that most of the computational savings come from performing fewer Monte Carlo simulations; skipping grid points nearly doubles the computational savings. As mentioned earlier, the advantage of the $\mathcal{GP}$ algorithm could be greater on a finer grid.

\section{Discussion}
The proposed algorithm significantly accelerates the Feldman-Cousins approach wherein experiments have to devote enormous computational resources in order to estimate uncertainties in neutrino oscillation parameters \cite{sousa2019aa}. This could also prove useful in estimating confidence intervals from a combined fit of neutrino oscillation results from different experiments when the respective likelihood functions are available. While we design the $\mathcal{GP}$ based construction in the neutrino oscillation context, the $\mathcal{GP}$ approximation does not have a particular parametric form. The same idea can therefore be applied to many other scenarios where the confidence interval construction for a continuous parameter over a bounded region normally proceeds via the unified approach. 

\bibliography{sample}
\bibliographystyle{unsrt}

\end{document}